\documentclass[final,5p,times,twocolumn]{elsarticle}

\usepackage{amsmath,amssymb,amsfonts}
\usepackage{graphicx}
\usepackage{textcomp}
\usepackage{lscape}
\usepackage{float}
\usepackage{xcolor}
\usepackage{orcidlink}
\usepackage{longtable}

\usepackage{booktabs}
\usepackage{rotating}
\usepackage{listings}
\usepackage{algpseudocode}
\usepackage{booktabs}
\usepackage[linesnumbered,ruled,vlined]{algorithm2e} 
\usepackage{hyperref}
\def\BibTeX{{\rm B\kern-.05em{\sc i\kern-.025em b}\kern-.08em
    T\kern-.1667em\lower.7ex\hbox{E}\kern-.125emX}}

\usepackage{tabularx,booktabs}
\usepackage[table]{xcolor}
\usepackage{threeparttable}
\usepackage{tikz}
\usepackage{array} 
\usepackage{multirow}

\newcolumntype{L}[1]{>{\raggedright\arraybackslash}p{#1}}
\definecolor{grey1}{HTML}{f5f5f5}

\newcolumntype{C}{>{\centering\arraybackslash}X} 
\journal{Journal}

\begin{document}

\begin{frontmatter}



\title{MeLeMaD: Adaptive Malware Detection via Chunk-wise Feature Selection and Meta-Learning}


\author[label1]{Ajvad Haneef K \corref{cor1} \orcidlink{0000-0002-2613-7131}}
 \cortext[cor1]{Corresponding author}
\ead{ajvad_p210054cs@nitc.ac.in}
\author[label1]{Karan Kuwar Singh}
\ead{karan_m220284cs@nitc.ac.in}
\author[label1]{Madhu Kumar S D\orcidlink{0000-0002-5276-8842}}
\ead{madhu@nitc.ac.in}

\affiliation[label1]{organization={Department of Computer Science and Engineering},
            addressline={National Institute of Technology Calicut}, 
            country={India}}

\begin{abstract}
    Confronting the substantial challenges of malware detection in cybersecurity necessitates solutions that are both robust and adaptable to the ever-evolving threat environment. The paper introduces Meta Learning Malware Detection (MeLeMaD), a novel framework leveraging the adaptability and generalization capabilities of Model-Agnostic Meta-Learning (MAML) for malware detection. MeLeMaD incorporates a novel feature selection technique, Chunk-wise Feature Selection based on Gradient Boosting (CFSGB), tailored for handling large-scale, high-dimensional malware datasets, significantly enhancing the detection efficiency. Two benchmark malware datasets (CIC-AndMal2020 and BODMAS) and a custom dataset (EMBOD) were used for rigorously validating the MeLeMaD, achieving a remarkable performance in terms of key evaluation measures, including accuracy, precision, recall, F1-score, MCC, and AUC. With accuracies of 98.04\% on CIC-AndMal2020 and 99.97\% on BODMAS, MeLeMaD outperforms the state-of-the-art approaches. The custom dataset, EMBOD, also achieves a commendable accuracy of 97.85\%. The results underscore the MeLeMaD's potential to address the challenges of robustness, adaptability, and large-scale, high-dimensional datasets in malware detection, paving the way for more effective and efficient cybersecurity solutions.
\end{abstract}




\end{frontmatter}


\section{Introduction}
\label{sec:introduction}
Malware poses an ever-increasing threat designed to wreak havoc on digital devices \citep{gorment2023machine}. Ranging from trojans to ransomware, these malicious entities target both individual computers and networks, resulting in heightened security vulnerabilities. The ramifications of cyberattacks continue to escalate: in 2023, cybercrime cost companies an estimated 8 trillion US dollars, a figure projected to soar to nearly 24 trillion US dollars by 2027 \citep{2024must79:online}. In parallel, the global indicator 'Estimated cost of cybercrime' is expected to increase by 6.4 trillion US dollars (a rise of 69.41\%) between 2024 and 2029, culminating in 15.63 trillion US dollars in 2029 and marking a new peak \citep{Globalcy47:online}. Given that more than one million new malware instances are generated daily, the urgency for robust and adaptive defense mechanisms against such threats has become paramount. Malware attacks have far-reaching consequences that affect individuals, businesses, and critical infrastructure. These attacks can compromise user identities, commit fraud, cause network downtime, breach sensitive data, violate privacy, disrupt services, erode trust, damage brand reputation, and lead to potential legal consequences \citep{liu2022deep}. Beyond these impacts, malware poses direct threats to user security by stealing private and account information, sending malicious text messages or emails, and even destroying the device's operating system, often without the user's knowledge \citep{dong2024android}. To counter such sophisticated and pervasive threats, security firms are increasingly leveraging advanced machine learning (ML) and deep learning (DL) methodologies to effectively detect and mitigate malware \citep{gopinath2023comprehensive,trizna2024nebula,deldar2023deep}. Malware is a universal threat capable of targeting diverse platforms, including Windows, Android, iOS, macOS, and Internet of Things (IoT) devices \citep{gopinath2023comprehensive, mafakheri2024android}.  As a result, detection techniques have been tailored and proposed for specific platforms \citep{jeon2024static,fang2023comprehensive}. 

Malware detection aims to classify programs or binary executable files as malicious or benign, often formulated as a binary classification problem \citep{gorment2023machine}. Machine learning (ML) plays a crucial role in this domain, offering advanced capabilities to identify and combat malicious software. ML models excel at summarizing discriminative features between malware and goodware, enabling the detection of both malware variants and zero-day samples \citep{maiorca2019towards}. Numerous ML and deep learning (DL)-based techniques have been proposed, demonstrating promising results in terms of accuracy and efficiency. These approaches have been widely adopted in cybersecurity, contributing to advances in malware detection methodologies.  
ML-based malware detection faces numerous challenges and limitations \citep{gorment2023machine}. However, a significant issue is the lack of high-quality, diverse, and representative datasets. Many existing datasets are outdated or biased, which hinders their generalization across unseen threats. The absence of standardized benchmarks further complicates the consistent evaluation of different models. The evolving nature of malware, characterized by obfuscation techniques such as polymorphism, metamorphism, and encryption, remains a substantial hurdle because these tactics often render static analysis and even some dynamic methods ineffective. Adversarial attacks present an additional challenge, allowing attackers to manipulate inputs to evade detection using ML models. The selection and extraction of relevant features are also critical, as the success of ML models is heavily dependent on the quality of the features used. Irrelevant or redundant features can lead to overfitting, reducing detection accuracy, while platform-specific malware behavior necessitates custom feature engineering, complicating cross-platform adaptability. Moreover, scalability and efficiency pose significant concerns. Dynamic and hybrid analysis methods, although more comprehensive, are resource-intensive and often unsuitable for real-time detection at scale. This issue is particularly acute for mobile and IoT devices, where limited computational resources make deploying complex ML models challenging. Finally, the interpretability and explainability of ML-based models remain critical limitations. Many ML and DL models function as black boxes, rendering their decision-making processes opaque. This lack of transparency poses challenges for trust, regulatory compliance, and the adoption of AI-driven systems in cybersecurity.

To address these challenges, this research makes the following contributions.
\begin{itemize}
\item MeLeMaD, a novel, first-ever framework for malware detection based on Model-Agnostic Meta-Learning (MAML), leveraging its adaptability and generalization capabilities to enhance detection models.
\item A new feature selection technique, Chunk-wise Feature Selection based on Gradient Boosting (CFSGB), tailored for large-scale, high-dimensional malware datasets.
\item Introduce a custom dataset EMBOD, which combines two benchmark datasets (EMBER\footnote{https://github.com/elastic/ember} and BODMAS\footnote{https://github.com/whyisyoung/BODMAS}) ensuring temporal diversity and improved generalization of the malware detection models.
\item Comprehensive evaluation of MeLeMaD using key performance metrics, particularly addressing the absence of MCC and limited use of AUC in existing studies.
\item Validation of MeLeMaD on two benchmark datasets (CIC-AndMal2020\footnote{https://www.unb.ca/cic/datasets/andmal2020.html} and BODMAS), and the custom dataset (EMBOD) demonstrating its scalability and robustness across diverse datasets.
\end{itemize}

The rest of the paper is structured as follows: Section \ref{sec:related-works} reviews the background and related works, covers learning-based methods for malware detection, feature selection techniques, and discusses state-of-the-art methods for malware detection. Section \ref{sec:methodology} details the proposed methodology of the MeLeMaD framework, which includes data pre-processing, feature selection using CFSGB, and binary classification (detection) using MAML. Section \ref{sec:experiments} presents the experimental results and discussion, encompassing the experimental settings, evaluation criteria, and analysis of the results. Finally, Section \ref{sec:conclusion} concludes the paper, summarizing key findings and suggesting future research directions.
\section{Background and Related Works}
\label{sec:related-works}

\subsection{Malware Detection}

Malware encompasses a broad range of harmful programs such as viruses, worms, trojans, ransomware, spyware, etc. \citep{gorment2023machine}. Each type of malware is designed to achieve specific objectives, such as financial theft, data exfiltration, or system disruption, and their diversity necessitates sophisticated detection mechanisms. The detection of malware is a critical task that aims at identifying and neutralizing malicious entities before they can execute harmful payloads. Malware detection employs three primary types of analysis—static, dynamic, and hybrid—to extract features that help to identify malicious behavior effectively \citep{gopinath2023comprehensive,gorment2023machine,aboaoja2022malwareissues}. Static analysis examines the code, structure, and metadata of malware without executing it. Static analysis is computationally efficient and resource-friendly, making it suitable for rapid initial assessments. However, it struggles against obfuscation techniques, such as encryption and dynamic loading, which modify or conceal the code to evade detection \citep{amira2023survey}. In contrast, dynamic analysis focuses on observing malware behavior during execution in a controlled environment, such as a sandbox, simulator, or emulator. It involves capturing runtime features such as API and system call traces, file modifications, network communications, and memory usage, providing deeper insights into malicious behavior. Dynamic analysis is more robust against obfuscation, as it reveals the malware's actual execution patterns rather than relying on its static code representation. However, dynamic analysis is resource-intensive, requiring the creation of a simulated or virtual environment to execute malware \citep{deldar2023deep}. Additionally, it may fail to cover all execution paths, as some malware may alter its behavior upon detecting the controlled environment, limiting the scope of the analysis. Hybrid analysis integrates the strengths of static and dynamic techniques to provide a comprehensive approach to feature extraction. By leveraging the advantages of both static and dynamic approaches, hybrid analysis provides a powerful mechanism for identifying increasingly sophisticated and evasive malware. However, hybrid analysis often demands significant computational power and memory, making it less efficient than either of these techniques alone.

Over the years, various malware detection methods have been proposed to combat these threats. Signature-based detection, a traditional cornerstone, identifies malware by matching unique patterns or sequences of bytes known as signatures against a pre-existing database \citep{deldar2023deep}. While efficient for detecting known threats, this method struggles against zero-day malware and variants that employ concealment techniques, such as encryption, polymorphism, and obfuscation. Even periodic database updates fail to keep pace with the rapidly evolving malware landscape, leading to high false-negative rates. Behaviour-based detection addresses this limitation by analysing the runtime behaviours of applications in isolated environments, such as communication packets, API calls, and system calls \citep{aboaoja2022malwareissues}. This method is effective against novel malware, but can result in false positives if the scenarios are not comprehensively evaluated. Heuristic-based detection expands upon behaviour-based techniques by using rule-driven models to detect suspicious patterns or deviations from normal activities. This involves two steps: first, establishing a baseline of normal system behaviour, and second, identifying anomalies relative to this baseline. Although heuristic approaches can identify both known and unknown malware, their generalized nature often leads to higher false-positive rates. Specification-based detection relies on monitoring and analyzing application behaviour against predefined specifications, making it a precise and manual method \citep{gorment2023machine}. Unlike heuristic approaches, specification-based methods focus on reducing false positives by adhering to the detailed system specifications. However, their reliance on manual processing limits the scalability. 

Learning-based detection, which is a recent advancement, leverages machine learning, deep learning, and artificial intelligence techniques to detect malware. These methods learn patterns from large datasets, enabling the identification of known and unknown malware variants. Despite their effectiveness, learning-based methods face challenges such as feature selection, scalability, interpretability, and adversarial attacks, necessitating further research to enhance their capabilities. Learning-based malware detection frameworks typically follow a structured pipeline to effectively analyze and classify potential threats \citep{maniriho2024systematic}. The process begins with data collection, in which diverse datasets comprising malware and benign samples are gathered from repositories or real-world environments. These datasets often include static features (e.g. code structure, API calls, and permissions) and dynamic features (e.g. system calls, network behavior, and memory usage) extracted through static, dynamic, or hybrid analysis \citep{amira2023survey}. The next step involves feature extraction and preprocessing, in which raw data are transformed into meaningful representations, such as vector-based, graph-based, or image-based formats, suitable for ML models \citep{aboaoja2022malwareissues}. Feature selection techniques are often employed to reduce dimensionality, eliminating irrelevant or redundant features to enhance the efficiency and performance of the model.  Once the dataset is preprocessed, it is fed into the model training phase, where supervised, unsupervised, or semi-supervised models are trained to differentiate between malicious and benign samples. Common models include Decision Trees, Support Vector Machines (SVMs), Convolutional Neural Networks (CNNs), Recurrent Neural Networks (RNNs), Autoencoders (AEs) and their variants, Generative Adversarial Networks (GANs), Restricted Boltzmann Machines (RBMS), Deep Belief Networks (DBNs), Attention-based models, and Ensemble models, depending on the type of features and data structure \citep{deldar2023deep,gaber2024malware,li2025malware, vasan2025advanced}. In the case of DL models, sophisticated architectures such as Graph Neural Networks (GNNs), Graph Attention Network (GATs),  Graph Convolutional Networks (GCNs) or Graph Autoencoders (GAEs) may be used to handle graph-based or high-dimensional representations \citep{mafakheri2024android, bilot2024survey}. The trained model is then validated and tested on unseen data to assess its performance using key metrics such as accuracy, precision, recall, and F1-score \citep{amira2023survey}. The final stage involves deployment and real-time detection, where the model is integrated into cybersecurity systems to classify incoming files or behaviors as malicious or benign. Real-time pipelines often incorporate ensemble models or meta-learning frameworks to adapt to evolving threats dynamically. Feedback loops are established to continuously update the model with new data, ensuring its robustness against zero-day attacks and evolving malware tactics. This end-to-end pipeline provides a comprehensive approach to malware detection, leveraging the adaptability and scalability of ML and DL frameworks to address the dynamic cybersecurity landscape.

\subsection{Related Works}
\begin{figure}[!ht]
\centering
\begin{tikzpicture}[scale=0.8]

    \begin{scope}[fill opacity=0.8]
        \draw[fill=green!30, draw=green!50] (-1.5,1) circle (3);
        \draw[fill=yellow!30, draw=yellow!50] (1.5,1) circle (3);
        
        \node at (-2,5) {Android};
        \node at (2,5) {Windows};
        
        \node[align=left] at (-3.5,1.5) {
            \small \cite{rahali2020didroid}\\
            \small \cite{batouche2021comprehensive}\\
            \small \cite{musikawan2022enhanced}\\
            \small \cite{ullah2022trojandetector}\\
            \small \cite{tidjon2022reliable}\\
            \small \cite{hammood2023machine}\\
            \small \cite{islam2023android}\\
            \small \cite{maniriho2024memaldet}\\
            \small \cite{ghourabi2024attention}\\
            \small \cite{li2025malware} \\
            \small \cite{mafakheri2024android}

        };
        
        \node[align=left] at (3.5,1.5) {
            
            \small \cite{hao2022eii}\\
            \small \cite{hussain2022hierarchical}\\
            \small \cite{lu2022self}\\
            \small \cite{manikandaraja2023rapidrift}\\
            \small \cite{chopra2023energy}\\
            \small \cite{jia2023ermds}\\
            \small \cite{hai2023proposed}\\
            \small \cite{dener2023clustering}\\
            \small \cite{robinette2024case}\\
            \small \cite{bhardwaj2024overcoming}\\
            \small \cite{xiong2024modified}
        };
        
        \node[align=center, rotate=90] at (0.2,1.5) {
            \small \cite{al2022parallel}\\
            \small \cite{vasan2025advanced}
        };
    \end{scope}
\end{tikzpicture}
\caption{Venn diagram illustrating the grouping of reviewed studies based on the platforms they target. The left (green) circle represents Android, the right (yellow) circle represents Windows, and the overlapping area indicates studies applicable to both platforms.}
\label{fig:lit-grouping-venn}
\end{figure}
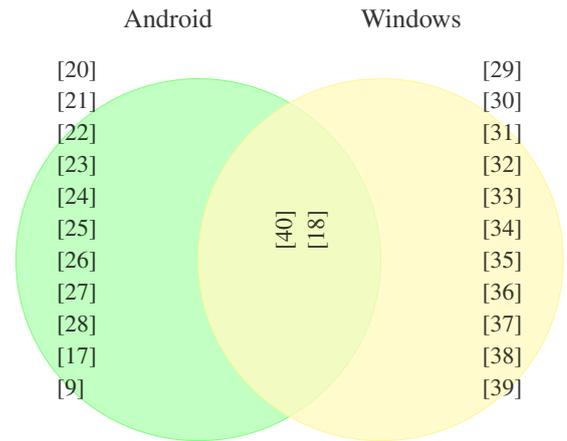

The field of malware detection has witnessed significant advancements through the application of various machine learning and deep learning models. Some recent methods have been depicted in Table \ref{tab:lit-review}. The table summarizes the dataset(s) used, methodologies employed, and limitations of each method and is organized chronologically to provide a clear overview of the evolution of research in this domain. The reviewed methods include only malware detection techniques proposed for Windows and Android platforms, as these are the most widely used operating systems. The platform-based grouping of reviewed methods shown in Figure \ref{fig:lit-grouping-venn}.

\begin{table*}[!ht]
    \centering
    \small
    \begin{threeparttable}
    \caption{Malware detection techniques based on  machine learning and deep learning}
    \label{tab:lit-review}
    \begin{tabular}{p{3cm}p{3cm}p{2.5cm}p{8cm}} 
        \toprule
        \textbf{Reference} &  \textbf{Dataset(s)}& \textbf{Method} &\textbf{Limitations} \\
        \midrule
        \cite{rahali2020didroid} & CIC-AndMal-2020 & CNN & Reliance on image representation increases computational cost \\  \midrule
        \cite{batouche2021comprehensive} & CIC-AndMal-2020 & RF & Limited by dataset size, lack of feature selection, and generalizability to real-world scenarios \\ \midrule
        \cite{al2022parallel} & BODMAS, CIC-AndMal-2020, DREBIN & PDL-FEMC & Higher computational costs for sequential implementations and scalability constraints due to inter-processor communication in distributed setups \\  \midrule
        \cite{musikawan2022enhanced} & CIC-InvesAndMal-2019, CIC-AndMal-2020, CIC-MalDroid-2020 & Ensemble ML + DNN & Computational complexity for large-scale, high-dimensional datasets may limit scalability \\ \midrule
        \cite{ullah2022trojandetector} & CIC-AndMal-2020, Contagio-Mobile, VirusShare & SVM, RF, LR, DT & Limited exploration of advanced deep learning methods, Lack of scalability for large, high-dimensional datasets \\ \midrule
        \cite{tidjon2022reliable} & CIC-MalMem-2022, CIC-AndMal-2020 & TDA + RF & Computationally intensive, limiting real-time applicability \\ \midrule
        \cite{hao2022eii} & Microsoft Malware Classification Challenge, BODMAS & CNN & Relies heavily on static analysis and may face challenges with obfuscated malware \\  \midrule
        \cite{hussain2022hierarchical} & BODMAS & RF, Ensemble ML & Relies on static features, which could be evaded by sophisticated malware obfuscation techniques \\  \midrule
        \cite{lu2022self} & BODMAS, BIG 2015 & Transformer-based models & May struggle with long-sequence modeling and very large datasets \\  \midrule
        \cite{islam2023android} & CIC-AndMal-2020 & Ensemble ML & Reliance on oversampling (SMOTE) may lead to overfitting and limits validation on unseen datasets \\  \midrule
        \cite{hammood2023machine} & CIC-AndMal-2020 & PSO + AGA & Complexity and real-time scalability remain challenges \\  \midrule
        \cite{manikandaraja2023rapidrift} & TRITIUM, INFERNO & Random Forest & Computational cost and restricted focus on Windows PE files, reducing generalizability to other platforms \\  \midrule
        \cite{chopra2023energy} & BODMAS & CNN + Transfer Learning & Limitations in handling obfuscated malware and requires further optimization for large-scale applications \\  \midrule
        \cite{jia2023ermds} & MalConv, EMBER & LB-MDS & Relies on synthetic obfuscation might not fully represent real-world threats \\  \midrule
        \cite{hai2023proposed} & BODMAS & EDR + CNN & Requires more validation for scalability \\  \midrule
        \cite{dener2023clustering} & BODMAS, EMBER, Kaggle & Clustering + CNN & High initial computational cost and potential overfitting in clustered subsets \\  \midrule
        \cite{robinette2024case} & BODMAS, Malimg & DNN & Provide verification of DNN models only, does not propose  a detectin framework  \\  \midrule
        \cite{bhardwaj2024overcoming} & BODMAS, Microsoft Kaggle & GANs + CNN & May face challenges with highly diverse datasets \\  \midrule
        \cite{maniriho2024memaldet} & MalMem-D2024 & DAE + Ensemble ML & Use only memory analysis dataset, static features were not considered
         \\  \midrule
         \cite{singh2024s} & CIC-AndMal-2020 & Attention + MLP & Relies on semantic mapping and dynamic analysis, detect ransomwares  \\  \midrule
        \cite{ghourabi2024attention} & CIC-AndMal-2020 & Attention + LightGBM   & Relies on static features, which could be evaded by sophisticated malware obfuscation techniques \\  \midrule
        \cite{xiong2024modified} & DataCon2020, BODMAS, EMBER & QCNN & Limited scalability \\  
        \bottomrule
    \end{tabular}
    \end{threeparttable}
\end{table*}

Figure \ref{fig:lit-grouping-venn} indicates that most studies focus on malware detection on a single platform rather than multiple platforms. Only a limited number of studies target both platforms. The studies encompass a wide range of techniques, including Convolutional Neural Networks (CNNs), Random Forest (RF), Support Vector Machines (SVM), and ensemble methods, and more sophisticated architectures such as Transformer-based models, Transfer Learning, and Generative Adversarial Networks (GANs), etc. The datasets utilized in these studies include well-known benchmarks such as CIC-AndMal-2020, EMBER, BODMAS, and DREBIN, which are essential for evaluating the performance of the proposed models. Few studies were evaluated using custom datasets introduced in their paper \citep{manikandaraja2023rapidrift,maniriho2024memaldet}. Most of the studies does not evaluate the frameworks based on on all key performance metrics such as accuracy, precision, recall, F1-score, Matthews Correlation Coefficient (MCC), and Area Under the Curve (AUC). The inclusion of these metrics is crucial for a comprehensive assessment of the models' effectiveness and reliability. Many models include only accuracy as evaluation metric, which may not provide a complete picture of the model's performance. The lack of inclusion of MCC in many existing studies is a significant oversight, as it is a valuable metric for evaluating binary classification models, especially in imbalanced datasets. The AUC is also often overlooked, despite its importance in assessing the model's ability to distinguish between classes across various thresholds. The absence of these metrics in many studies limits the comparability and generalizability of the proposed models. Irrespective of the platform targeted, most the papers reviewed consider the static features for malware detection. Some studies use image-based representations of malware samples and employ architectures such as CNNs, Transfer Learning etc. for classification \citep{hao2022eii,robinette2024case}. Ensemble methods are also prevalent, combining multiple classifiers to enhance detection accuracy and robustness \citep{musikawan2022enhanced,rahali2020didroid}. Attention mechanisms and Transformer-based models have gained traction in recent years, demonstrating their effectiveness in capturing complex patterns and relationships within malware data \citep{lu2022self,ghourabi2024attention}. Due to the computational complexity of large-scale, high-dimensional datasets, many studies fail to train the model on the whole dataset, instead a subset is used for training and evaluating the model.  Although only limited studies use specific feature selection techniques, in order to reduce the dimention of the feature space before trainining the models\citep{rahali2020didroid,al2022parallel,islam2023android}. Other studies that does not apply any feature selection techniques, which may lead to more complex models and increased computational costs \citep{hussain2022hierarchical,manikandaraja2023rapidrift}. The reliance on static features in many studies raises concerns about the models' robustness against sophisticated malware obfuscation techniques, which can evade detection by altering their code structure or behavior \citep{hussain2022hierarchical,ghourabi2024attention}. Few studies focus on detection of malware alone, while others also consider the classification of malware families \citep{rahali2020didroid,batouche2021comprehensive}. The reviewed studies highlight the potential of machine learning and deep learning techniques in addressing the challenges of malware detection. However, they also underscore the limitations and challenges faced in the field, such as computational complexity, reliance on static features, lack of efficient feature selection techniques, and challenges in real-time applicability. The need for larger datasets to enhance generalizability is also emphasized. The reviewed studies demonstrate the potential of machine learning and deep learning techniques in addressing the challenges of malware detection while also highlighting areas for improvement and future research directions.

The field of malware detection has made notable advancements by leveraging diverse machine learning (ML) and deep learning approaches. Studies have demonstrated the effectiveness of classifiers such as Random Forest, CNNs, and hybrid frameworks for both binary and multiclass malware detection. However, limitations persist across key areas, including handling high-dimensional datasets, scalability to real-time applications, and generalizability to unseen data. Feature selection techniques such as PSO and SMOTE have been employed to optimize performance, but their application is often dataset-specific and computationally intensive. Approaches such as AMDI-Droid and MD-ADA highlight the potential of hybrid and adversarial domain adaptation techniques, yet their reliance on labeled data and high computational overhead limits scalability. Similarly, innovative frameworks sucha s EII-MBS and SeqConvAttn excel in accuracy but struggle with obfuscated malware and real-time applicability. Several studies underscore the challenges posed by evolving malware threats, concept drift, and the need for robust feature representations. Open issues include improving robustness against adversarial attacks, enhancing efficiency for large-scale deployments, and exploring advanced feature selection and regularization techniques to mitigate overfitting and computational constraints. These challenges highlight the need for future research to prioritize scalability, cross-platform applicability, and adaptive methodologies for evolving threat landscapes.

\section{Proposed Methodology}
\label{sec:methodology}
\begin{figure*}[!ht]
\centering
\includegraphics[width=0.9\textwidth]{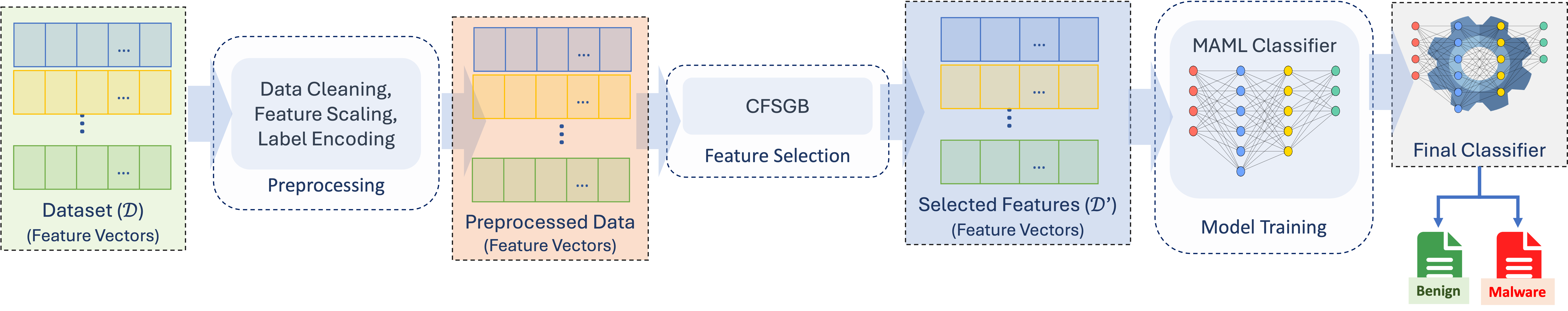} 
\caption{Architecture of MeLeMaD Framework (Proposed)}
\label{fig:framework}
\end{figure*}

The methodology section provides an overview of the proposed MeLeMaD framework, consists of several key components, including data preprocessing, feature extraction, feature selection, model training, and evaluation strategies.  The MeLeMaD framework is built for malware detasets, has been experimented on two benchmark datasets, CIC-AndMal2020\citep{rahali2020didroid} and BODMAS\citep{yand2021bodmas}, as well as a custom dataset named EMBOD (a combination of two benchmark datasets: EMBER and BODMAS). While the core methodology remains consistent, certain steps, particularly feature extraction and data preprocessing, were tailored to the unique characteristics of each dataset. These dataset-specific details are discussed in the Experiments section (Section \ref{sec:experiments}). The framework is designed to leverage the Gradient Boosting and Model-Agnostic Meta-Learning (MAML) to enhance the adaptability and generalization capabilities of malware detection models. The architecture of the MeLeMaD framework is illustrated in Figure \ref{fig:framework}. The framework is designed to be flexible, scalable, and efficient, enabling rapid learning from limited data and quick adaptation to new tasks. The key steps involved in the MeLeMaD framework are described in detail below. The variables used in the MeLeMaD framework are summarized in Table \ref{tab:notation}.

\begin{table}[!htbp]
    \centering
    \begin{threeparttable}
    \caption{Variables used in MeLeMaD Framework}
    \label{tab:notation}
    \begin{tabular}{c L{6cm}}
        \toprule
        \textbf{Variable} &  \textbf{Description} \\
        \midrule
        $\mathcal{D}$ & Initial dataset \\ \midrule
        $\mathcal{D}'$ & Dataset based on selected global features \\ \midrule
        $\mathcal{C}_i$ & Chunk $i$ of the dataset \\ \midrule
        $l$ & Size of each chunk (number of samples) \\ \midrule
        $o$ & Overlap size between consecutive chunks \\ \midrule
        $p$ & Chunk size as a fraction of the total dataset \\ \midrule
        $q$ & Overlap size as a fraction of the total dataset \\ \midrule
        $k$ & Total number of chunks \\ \midrule
        \(f^{i}_j\) & Important of score of feature $j$ in chunk $i$ \\ \midrule
        $\mathcal{FI}_i$ & Feature importance vector for chunk $i$ \\ \midrule
        $\mathcal{T}$ & Predefined threshold for feature selection \\ \midrule
        $\mathcal{S}_i$ & Selected features in chunk $i$ \\ \midrule
        $\mathcal{S}_{global}$ & Global set of selected features across all chunks \\ \midrule
       $\mathcal{D}'_{train}$ & Meta-training set \\ \midrule
        $\mathcal{D}'_{test}$ & Meta-testing set \\ \midrule
        ${\mathcal{T}_i}$ & Task $i$ in the meta-learning network \\ \midrule
        $\mathcal{D}_{task}$ & Dataset for task $i$  \\ \midrule
        $\mathcal{P}_i$ & Support set for  task $i$  (subset of $\mathcal{D}_{task}$) \\ \midrule
        $\mathcal{Q}_i$ & Query set for  task $i$  (subset of $\mathcal{D}_{task}$) \\ \midrule
        $\mathcal{K}$ & Number of classes \\ \midrule
        $\mathcal{M}_{\theta}$ & Meta-learning base model with parameters $\theta$ \\ \midrule
        \(\theta'_i\) & Adapted parameters for task $i$ \\ \midrule
        $\mathcal{L}_{\text{support}}\bigl(\mathcal{P}_i\bigr)$ & Loss function on the support set for  task task $i$ \\ \midrule
        $\mathcal{L}_{\text{query}}\bigl(\mathcal{Q}_i\bigr)$ & Loss function on the query set for task task $i$ \\ \midrule
        $\mathcal{L}_{\text{meta}}$ & Meta-loss function for updating the model parameters \\ \midrule
        $\nabla_\theta$  & Gradient of the support loss function with respect to the model parameters \\ \midrule
        $\nabla_{\theta_{\text{meta}}}$ & Gradient of the meta-loss function with respect to the model parameters \\ 
        \bottomrule
    \end{tabular}
    \end{threeparttable}
\end{table}

The malware detection problem can be formulated as a binary classification task, where the goal is to classify samples into two categories: malware and benign. Consider a dataset $\mathcal{D} = \{(x_i, y_i)\}_{i=1}^n$, where $n$ represents the total number of samples in the dataset. Each sample $x_i \in \mathbb{R}^m$ is a feature vector of dimension $m$, extracted from either static, dynamic, or hybrid analysis of the sample. Each $y_i \in \{0, 1\}$ is a binary label, where $y_i = 1$ indicates that the sample is malware and $y_i = 0$ indicates that the sample is benign.
The malware detection model can be defined as a function:
\begin{equation}
    f: \mathbb{R}^m \to \{0, 1\}
\end{equation}

The goal of the malware detection model is to learn a mapping from the feature space to a binary output \(\hat{y}_i\), where \(\hat{y}_i = 1\) predicts that \( x_i \) is malware, and \(\hat{y}_i = 0\) predicts that \( x_i \) is benign. The model output for a given sample \( x_i \) is given by:
\begin{equation}
\hat{y}_i = f(x_i).
\end{equation}

\subsection{Feature Extraction and Data Preprocessing}
The feature extraction and data preprocessing, were tailored to the unique characteristics of each dataset. These dataset-specific details are discussed in the Experiments Section (Section \ref{sec:experiments}).
\subsection{Feature Selection using CFSGB}
\begin{figure*}[h!]
     \centering
    \includegraphics[width=\textwidth]{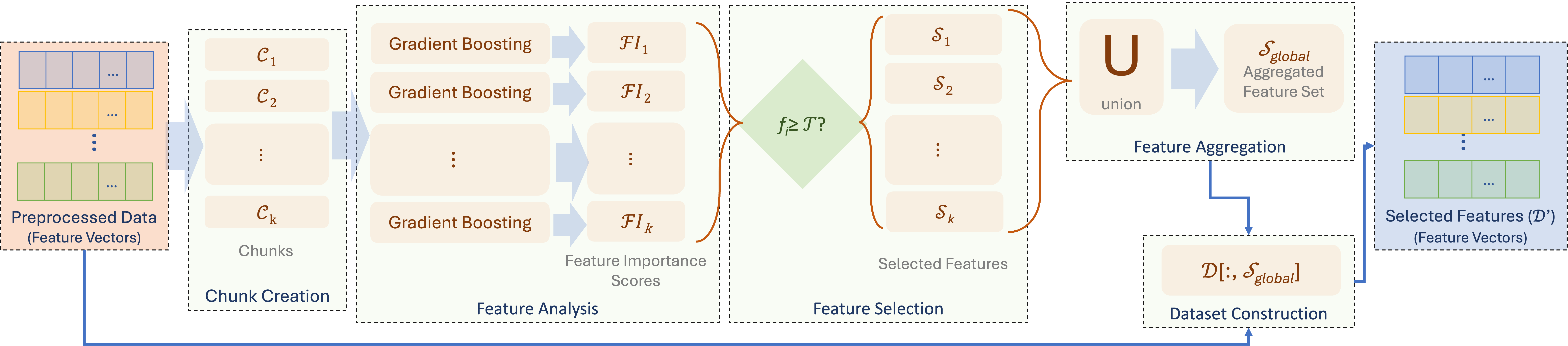}   
        \caption{Architecture of Chunk-wise Feature Selection based on Gradient Boosting (CFSGB)}
        \label{fig:block-diagram-cfsgb}
    \end{figure*}

For feature selection, a noval feature selection technique Chunk-wise Feature Selection based on Gradient Boosting (CFSGB) has been proposed in this paper. The malware datasets typically consist of a large number of features,
necessitating the identification of relevant attributes to optimize
model training and testing. In addition the size of the dataset
is very large, which demands substantial memory and powerful
hardware resources. The state-of-the-art feature selection techniques are not efficient in handling such large-scale, high-dimentional datasets. Most of
the feature selection techniques work on the entire dataset, which
is computationally expensive and requires significant memory
resources. To address these challenges, we propose a novel
feature selection technique, Chunk-wise Feature Selection based
on Gradient Boosting (CFSGB), that partitions the dataset into
overlapping chunks and evaluates feature importance within
each segment. The CFSGB leverages gradient boosting, a robust
machine learning algorithm known for its ability to handle complex data relationships and provide accurate feature importance
metrics. It identifies and prioritizes features that significantly
contribute to malware detection, thereby streamlining the feature
selection process and enhancing model efficiency. The CFSGB aims to optimize feature selection in malware detection by leveraging gradient
boosting on segmented data chunks. The architecture of the
CFSGB is shown in Figure \ref{fig:block-diagram-cfsgb} and comprises the following
steps:
\subsubsection{Chunk Creation}
The first step involves dividing the dataset into manageable, overlapping chunks to facilitate comprehensive feature selection. The dataset \(\mathcal{D}=\{X,y\}\) comprises \(n\) instances, where $X \in \mathbb{R}^{n \times m}$ represents the feature matrix with $m$ features, and $y \in \mathbb{R}^{n}$ represents the target labels (malware or benign files). The chunk creation process generates \(k\) overlapping chunks, denoted as \[\mathcal{C}_1, \mathcal{C}_2, \dots, \mathcal{C}_k\] with the following characteristics:
\begin{itemize}
    \item Each chunk $\mathcal{C}_i \in \mathbb{R}^{l \times m}$, contains $l = p \times n$ samples, where $p$ is the chunk size as a fraction of the total dataset (e.g., $p = 0.2$ indicates that each chunk comprises 20\% of the dataset).
    \item Consecutive chunks overlap by $o = q \times n$ samples, where $q$ is the overlap size as a fraction of the total dataset (e.g., $q = 0.1$ indicates that each chunk overlaps by 10\%).
\end{itemize}
In addition, the number of chunks $k$ is calculated as:
$k = \left\lfloor \frac{n}{l} \right\rfloor$

The overlapping of samples in between chunks ensures that each feature has the opportunity to influence multiple segments of the analysis, thereby enhancing the robustness and reliability of feature selection.

\subsubsection{Feature Analysis using Gradient Boosting}
Once the dataset has been divided into overlapping chunks, the next step involves applying a gradient-boosting classifier to each segment to evaluate feature importance. Gradient boosting is a powerful ensemble learning technique that builds a series of decision trees to predict the target variable. It is known for its ability to handle complex data relationships and provide accurate feature importance metrics. For each chunk $\mathcal{C}_i$, the gradient-boosting classifier is trained using the features $X_i$ and target labels $y_i$ specific to that segment. The model calculates the importance scores for each feature, which gauge how much each feature influences the classifier's predictive accuracy within its respective chunk. Features with importance scores exceeding zero are prioritized because they demonstrate meaningful contributions toward identifying the target feature. This approach ensures that only features making significant contributions to the classifier's predictive capabilities are retained. By focusing on features with nonzero importance scores, we streamline the dataset to include only the most influential attributes for subsequent analyses or model training tasks.

For each chunk $\mathcal{C}_i$, by applying gradient boosting, we obtain the feature importance scores \(\mathcal{FI}_i = \{f^{i}_1, f^{i}_2, \dots, f^{i}_m\}\), where \(f^{i}_j\) represents the importance score of the feature \(j\) in chunk $i$. By analyzing each chunk independently, we obtain localized importance scores that capture the relevance of features within specific data segments. This localized approach ensures that feature selection is sensitive to variations across the dataset.

\subsubsection{Chunkwise Feature Selection Based on Localized Importance Scores}
After obtaining the feature importance scores for each chunk $\mathcal{C}_i$, we perform feature selection based on a predefined threshold ($\mathcal{T}$). Features with importance scores exceeding the threshold are considered significant and retained for further analysis, while those below the threshold are discarded. The threshold value is determined empirically based on the distribution of importance scores across all chunks. 
In general, for each chunk $\mathcal{C}_i$, the selected feature set  $\mathcal{S}_i$ is denoted as:
\[
\mathcal{S}_i = \{f^{i}_j \mid f^{i}_j \geq \mathcal{T}\}
\]
By setting a consistent threshold ($\mathcal{T}$) for all chunks, we ensure that the feature selection process is standardized and robust across the dataset. This approach enables the identification of features that consistently contribute to predictive accuracy and model performance, enhancing the reliability and interpretability of the selected attributes.

\subsubsection{Feature Aggregation}
The next step involves aggregating the selected features from each chunk to create a consolidated feature set that captures the most relevant attributes across the dataset. Since different chunks may select different sets of features, the global feature set $\mathcal{S}_{global}$ is formed by taking the union of all the selected features across the chunks:
\[
\mathcal{S}_{global} = \bigcup_{i=1}^{k} \mathcal{S}_i
\]
The aggregated feature set $\mathcal{S}_{global}$ contains all the features that are deemed important in at least one chunk, ensuring that no relevant feature is excluded from the final analysis. By combining the selected features from each segment, we create a comprehensive feature set optimized for malware detection and model training tasks. This consolidated approach enhances the robustness and effectiveness of the feature selection process, enabling the identification of critical attributes that contribute to accurate malware detection.

\subsubsection{Dataset Construction with Selected Global Features}
Based on the aggregated feature set $\mathcal{S}_{global}=\{f_1,f_2,\dots, f_r\}$ , where $r$ is the number of selected features ($r<m$), we construct a new dataset $\mathcal{D}'=\{X',y\}$ that includes only the selected features $X'$ and original target labels $y$. The new feature matrix $X' \in \mathbb{R}^{n \times r}$ is formed by retaining the instances from the original dataset $\mathcal{D}$ and selecting the features in $\mathcal{S}$. The new dataset $\mathcal{D}'$ can be defined as:
\begin{equation}
\mathcal{D}' = \{X',y\}
\end{equation}
where the new feature matrix $X'$ is given by:
\begin{equation}
X' =  \{X_j\mid j \in \mathcal{S}\}
\end{equation}
This streamlined dataset $\mathcal{D'}$ is optimized for subsequent analyses or model training tasks, focusing on the most relevant attributes identified through the feature selection process. By tailoring the dataset to the specific requirements of malware detection, we enhance the efficiency and effectiveness of the feature matrix, enabling more accurate and interpretable analyses. The new dataset is designed to contain only the most informative features, ensuring that the subsequent model training tasks are optimized for malware detection. By focusing on the most relevant attributes identified through the feature selection process, we enhance the efficiency and effectiveness of the dataset, enabling more accurate and interpretable analyses. The new dataset is tailored to the specific requirements of malware detection, ensuring that only the most informative features are retained for further processing.

The steps for the entire process of CFSGB is depicted in Algorithm \ref{algorithm:cfsgb}. The algorithm outlines the steps involved in chunk creation, feature analysis, aggregation, and dataset construction, culminating in the creation of a streamlined dataset optimized for malware detection. By systematically evaluating feature importance across overlapping data segments, CFSGB enhances the efficiency and accuracy of feature selection, enabling more effective model training and malware detection.

\begin{algorithm}[h]
    \caption{Chunk-wise Feature Selection based on Gradient Boosting (CFSGB)} \label{algorithm:cfsgb}
    \KwIn{Dataset $\mathcal{D} = \{X, y\}$ with $n$ samples and $m$ features}
    \KwOut{Selected feature set $\mathcal{S}_{\text{global}}$ and new dataset $\mathcal{D'}$}
    
    \SetKwFunction{TrainModel}{TrainModel}
    \SetKwFunction{ComputeFeatureImportance}{ComputeFeatureImportance}
    \SetKwFunction{SelectFeatures}{SelectFeatures}
    \SetKwFunction{Aggregate}{Aggregate}
    \SetKwFunction{ReconstructDataset}{ReconstructDataset}
    \SetKwFunction{TrainFinalModel}{TrainFinalModel}
    
    \BlankLine
    
    \textcolor{blue}{\underline{\textbf{Step 1: Chunk Creation}}} \\
    Divide dataset $\mathcal{D} = \{X, y\}$ into $k$ overlapping chunks $\{\mathcal{C}_1, \mathcal{C}_2, \dots, \mathcal{C}_k\}$, where each chunk $\mathcal{C}_i$ contains $p \times n$ samples, with $o \times n$ overlap between consecutive chunks\;
     
    \For{$i = 1$ to $k$}{
        \textcolor{blue}{\underline{\textbf{Step 2: Feature Analysis}}} \\
        Train a gradient boosting model on chunk $\mathcal{C}_i$\;
        Compute feature importance scores $\mathcal{FI}_i = \{f_1^i, f_2^i, \dots, f_m^i\}$ for chunk $\mathcal{C}_i$\;
        
        \textcolor{blue}{\underline{\textbf{Step 3: Feature Selection}}} \\
        Select feature set $\mathcal{S}_i$ from $\mathcal{C}_i$ based on the threshold $\tau$:
        \[
        \mathcal{S}_i = \{f_j^i \mid f_j^i > \tau\}
        \]
    }
    
    \textcolor{blue}{\underline{\textbf{Step 4: Feature Aggregation}}} \\
    Aggregate the selected feature sets $\{\mathcal{S}_1, \mathcal{S}_2, \dots, \mathcal{S}_k\}$ to form the global feature set $\mathcal{S}_{\text{global}}$:
    \[
    \mathcal{S}_{\text{global}} = \bigcup_{i=1}^{k} \mathcal{S}_i
    \]
    
    \textcolor{blue}{\underline{\textbf{Step 5: Dataset Construction with Selected Features}}} \\
    Reconstruct the dataset using the selected features $\mathcal{S}_{\text{global}}$:
    \[
    X' = \{X_j \mid j \in \mathcal{S}_{\text{global}}\}
    \]
    Form the new dataset $\mathcal{D'} = \{X', y\}$\;

    \Return $\mathcal{S}_{\text{global}}$, $\mathcal{D}'$\;

    \end{algorithm}

\subsection{Classification using Model Agnostic Meta Learning}
\begin{figure*}[!ht]
\centering
\includegraphics[width=0.9\textwidth]{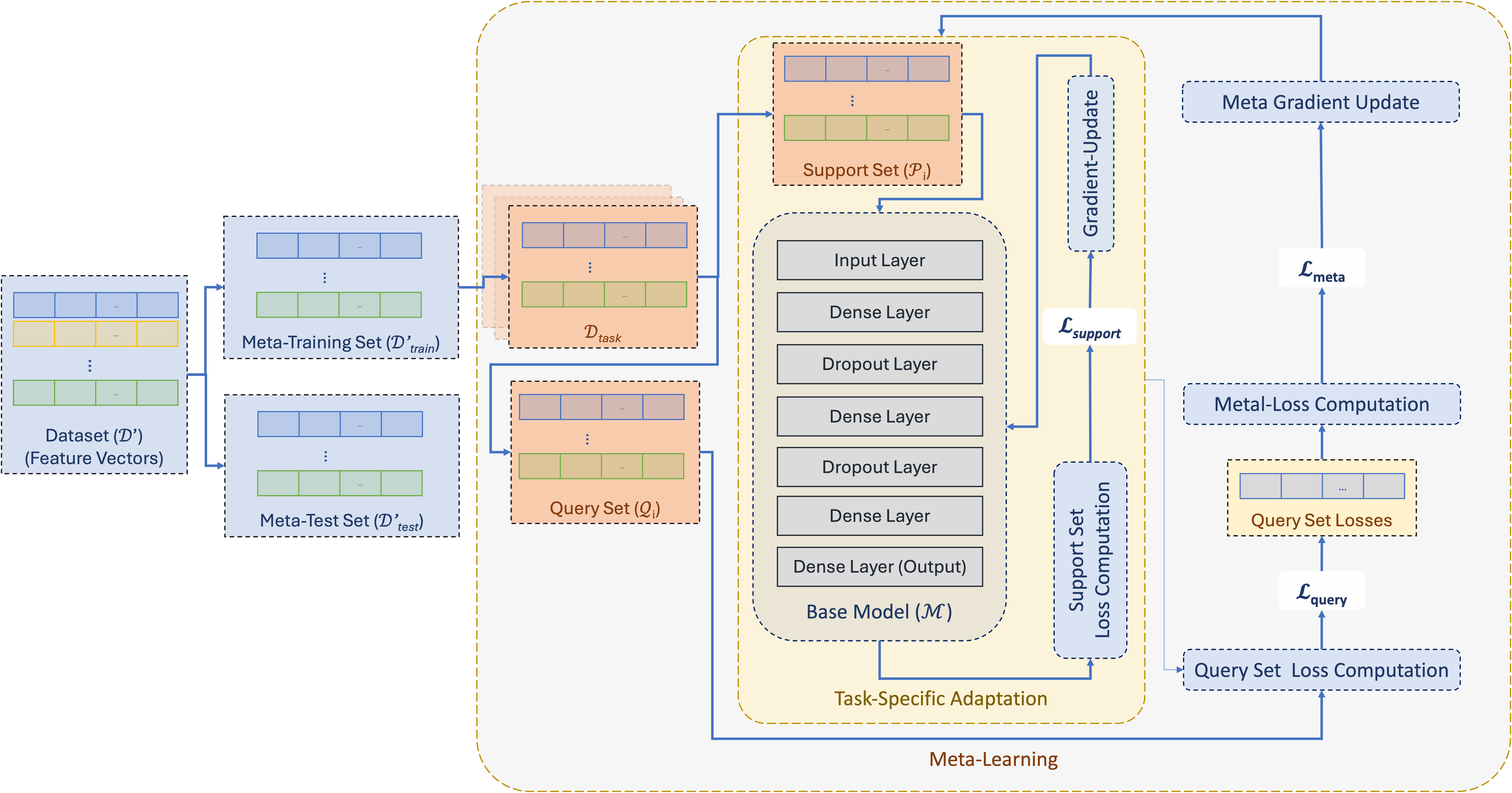} 
\caption{MAML training process}
\label{fig:maml-training}
\end{figure*}

MeLeMaD leverages the Model Agnostic Meta Learning (MAML) to build the classification model for enhanced adaptability and generalization capabilities. Model Agnostic Meta Learning(MAML) is a sophisticated machine learning approach designed to enhance the adaptability and generalization capabilities of models across diverse tasks \citep{finn2017model,tak2024enhancing}. The fundamental objective of MAML is to enable a model to quickly adapt to new tasks with minimal data by learning a set of broadly applicable initial parameters. This meta-learning paradigm is referred to as ”model-agnostic” because MAML is applicable to various types of models, irrespective of their architectures or specific learning algorithms. The key idea behind MAML is to train a model’s parameters in such a way that, after a few gradient updates on a new task (task-specific adaptation), the model achieves optimal performance. The training process involves two nested loops: an outer loop and an inner loop. In the outer loop, the model is updated to be able to quickly adapt to new tasks, while the inner loop involves rapid adaptation to specific tasks. The training process of the MAML classifer has been depicted in Figure \ref{fig:maml-training}.

The keys steps involved in the learning process has been described as follows:
\begin{enumerate}
    \item \emph{Initialization}: The model is initialized with parameters that are generally suitable for a wide range of tasks.
    \item \emph{Outer Loop (Meta-Training)}: The model is exposed to a variety of tasks during the meta-training phase. For each task, the model’s parameters are adjusted using a small set of training data to improve its ability to adapt quickly. The objective is to optimize the model’s parameters in a way that facilitates efficient learning across diverse tasks.
    \item \emph{Inner Loop (Task-Specific Adaptation)}: For each task, a subset of the training data is used to update the model’s parameters through gradient descent.This process allows the model to rapidly adapt to the specifics of each task.
    \item \emph{Meta-Testing (Evaluation)}: After the meta-training phase, the model’s ability to generalize and adapt quickly is evaluated on new tasks or test data.
\end{enumerate}

MAML’s strength lies in its ability to learn a set of initial parameters that serve as a strong foundation
for quick adaptation to new tasks. This makes it particularly useful in scenarios where obtaining a large amount of labeled data for each specific task is impractical or expensive.

The training process of the MAML classification model involves four steps: dataset splitting, base model initialization, task formation, and model training.
\subsubsection{Dataset Splitting and Sampling}
The new dataset $\mathcal{D}' = \{X',y\}$ obtained after feature selection by applying CFSGB, is subdivided into two subsets: the Meta-Training Set ($\mathcal{D}'_{train}$) and Meta-Test Set ($\mathcal{D}'_{test}$). The Meta-Training subset ($\mathcal{D}'_{train}$) constitutes the core of the MAML training process. While the Meta-Test subset ($\mathcal{D}'_{test}$) serves as the final evaluation benchmark, used after the meta-training process is completed.

\subsubsection{Base Model Initialization}
The base model ($\mathcal{M}_{\theta}$) is designed as deep neural network parameterized by \(\theta\), which may be initialized with random or pretrained parameters. Throughout the MAML process, these parameters \(\theta\) are iteratively updated to make the model quickly adaptable to new tasks. The base model architecture comprises an input layer, followed by a dense (fully connected) hidden layer, a dropout layer for regularization and subsequently, two additional dense layers, concluding with an output layer for binary classification, as indicated in Figure \ref{fig:maml-training}. This lightweight architecture ensures fast adaptation during few-shot learning, while maintaining sufficient capacity to capture the underlying patterns in malware features across various tasks.
\subsubsection{Task Formation} 
The MAML network is made to work with few-shot learning tasks (${\mathcal{T}_i}$). Each task ($\mathcal{T}_i$) consumes a small subset ($\mathcal{D}_{task}$) selected from the meta-training set ($\mathcal{D}'_{train}$) to classify them based on extracted features. This task-specific subset ($\mathcal{D}_{task}$) is further split into two sets, support set ($\mathcal{P}_i$) and query set  ($\mathcal{Q}_i$) , used to train and validate the few short learners respectively. 
The objective  of the few-shot learning task ($\mathcal{T}_i$) is to correctly classify the query set ($\mathcal{P}_i$) by learning from the support set ($\mathcal{Q}_i$). Typically, the few shot classification is made to split the dataset over classes, rather than over samples. Also, each few shot task defined over different classes. Since, the binary classification is used in the few-shot learning task, we define the number of classes as two ($\mathcal{K}=2$).
For instance, in context of malware detection (binary classification), each task is defined as classifying the samples in $\mathcal{D}_{task}$ as maliciuos or benign. The few shot base models learn to make accurate predictions with very limited data by finetuning on this task-specific dataset.

\subsubsection{Model Training} 
The MAML classiffier training is a two step approach, consising of an inner loop and outer loop. Both, in together enables the fast adaptation of the trained model. The innerloop is used for task-specific learning in which the few short learners are trained and fine-tuned using the support set and  and evaluated using the query set. While, the outer loop aggregates the few-shot tasks and updates the initial parameters to improve the generalizability of the final model.    
\begin{itemize}
    \item \textit{Inner Loop (Task-Specific Learning)}: In inner loop, each few shot task is defined as the classification of malware samples by learning from a few-shot training batch or support set. The base models (few-shot learners) with radomly initilized parameters, are finetuned by computing the gradient of their loss. The model’s parameters are updated using these gradients through gradient descent. This updated model is now slightly adapted to the specific task.
    Formally, if a base model $\mathcal{M}_{\theta}$ has parameters \(\theta\), the support set is used to adapt \(\theta \rightarrow \theta'_i\) via a few gradient steps on the support loss
    \(\mathcal{L}_{\text{support}}(\mathcal{P}_i)\), such that:
    \begin{equation}
    \theta'_i \;=\; \theta \;-\; \alpha \,\nabla_\theta \,\mathcal{L}_{\text{support}}(\mathcal{P}_i)
    \end{equation}
    where \(\alpha\) is the inner-loop learning rate. The support loss is computed by evaluating the model \(\mathcal{M}_{\theta}\) on the support set \(\mathcal{P}_i\):
    \begin{equation}
        \mathcal{L}_{\text{support}}\bigl(\mathcal{P}_i\bigr)\;=\; - \frac{1}{N} \sum_{i=1}^{N}\Bigl[l_i\log({p_i})+ (1-l_i) \log(1-p_i)\Bigr]
    \end{equation}

    where, \(N\) denotes the total number of samples in the support set (\(\mathcal{P}_i\)), \(l_i\) represents the ground-truth label for the \(i\)-th sample, and \(p_i\) is the predicted probability that the \(i\)-th sample is malicious.

    After the inner loop adaptation, the updated model parameters \(\theta'_i\) are evaluated on the query set \(\mathcal{Q}_i\) to assess how well the model generalizes to new samples after fine-tuning. The query loss is computed similarly to the support loss, but using the samples in \(\mathcal{Q}_i\). Formally, if the query set \(\mathcal{Q}_i\) contains \(M\) samples, the query loss is given by

    \begin{equation}
    \mathcal{L}_{\text{query}}\bigl(\mathcal{Q}_i\bigr) \;=\; - \frac{1}{M} \sum_{j=1}^{M}\Bigl[l_j \log(q_j) + (1-l_j) \log(1-q_j)\Bigr]
    \end{equation}

    where \(M\) denotes the total number of samples in the query set (\(\mathcal{Q}_i\)), \(l_j\) is the ground-truth label for the \(j\)-th sample in the query set, and \(q_j'\) is the predicted probability (using the updated parameters \(\theta'_i\)) that the \(j\)-th sample is malicious. This query loss, \(\mathcal{L}_{\text{query}}(\mathcal{Q}_i)\), is then used in the outer loop to aggregate gradients across different few-shot tasks, guiding the meta-learner to update the initial parameters \(\theta\) such that the model becomes rapidly adaptable to new tasks.

    This setup allows the model to learn from only a handful of examples in \(\mathcal{P}_i\) and generalize well to \(\mathcal{Q}_i\), which is critical in malware detection scenarios where benign and malicious samples are limited for each few-shot task.

    \item \textit{Outer Loop (Meta-Learning)}: The outer loop in the MAML network is responsible for learning a good initialization \(\theta_{\text{meta}}\) that enables the model to quickly adapt to new tasks with minimal fine-tuning. Unlike the inner loop, which adapts task-specific parameters using a small support set, the outer loop aggregates information across multiple few-shot tasks and updates the meta-learner to improve generalizability. After task-specific fine-tuning in the inner loop, each task produces a query loss \(\mathcal{L}_{\text{query}}\bigl(\mathcal{Q}_i\bigr)\), which quantifies how well the model generalizes after adaptation. The outer loop then aggregates these query losses across all sampled tasks and optimizes the meta-learner parameters using gradient descent.  
    
    Formally, for a batch of \(T\) tasks, the meta-loss is defined as the average of query losses over all tasks:
    
    \begin{equation}
    \mathcal{L}_{\text{meta}} 
    \;=\; \frac{1}{T}
    \sum_{i=1}^{T} \mathcal{L}_{\text{query}}\bigl(\mathcal{Q}_i\bigr)
    \end{equation}
    
    The gradients of the meta-loss with respect to the meta-learner parameters \(\theta_{\text{meta}}\) are then computed, and the parameters are updated as:
    
    \begin{equation}
    \theta_{\text{meta}} 
    \;\leftarrow\; 
    \theta_{\text{meta}} 
    - 
    \beta 
    \,\nabla_{\theta_{\text{meta}}} 
    \,\mathcal{L}_{\text{meta}}
    \end{equation}
    
    where \(\beta\) is the outer-loop learning rate. This update allows the meta-learner to refine its initialization so that future few-shot learners can learn more effectively from small amounts of data with fewer gradient steps. By iteratively refining \(\theta_{\text{meta}}\) over multiple tasks, the model gains the ability to generalize well across different malware classification scenarios, ensuring robust detection even in low-data conditions.
   

    \item \textit{Meta-Testing (Evaluation)}:
    In the Meta-Testing stage, the meta-learner—already trained on the Meta-Training set  is evaluated on an unseen set of tasks drawn from the Meta-Test dataset ($\mathcal{D}'_{test}$). This final evaluation step assesses the model’s ability to generalize and adapt quickly to new tasks, providing insights into its overall performance and effectiveness. The model’s performance on the meta-test set serves as a benchmark for assessing its generalizability and robustness in real-world scenarios. 
\end{itemize}
\section{Experiments}
\label{sec:experiments}
\subsection{Datasets}
We have conducted extensive experiments on two benchmark datasets, CIC-AndMal2020 and BODMAS, as well as a custom dataset named EMBOD, introduced in this research, which is a combination of two benchmark datasets: EMBER and BODMAS.

\subsubsection{CIC-AndMal-2020 Dataset}
\par The CIC-AndMal-2020\citep{rahali2020didroid} dataset is a comprehensive and large-scale Android malware dataset comprising a total of $\sim$400k Android applications, evenly split into $\sim$200k benign samples and $\sim$200k malware samples. The benign samples were sourced from the AndroZoo dataset, ensuring dataset balance. The malware samples were collected in collaboration with CCCS (Canadian Centre for Cyber Security), where each sample was labeled and categorized into corresponding malware families. The dataset encompasses 14 prominent malware categories, including Adware, Backdoor, File Infector, No Category, Potentially Unwanted Apps (PUA), Ransomware, Riskware, Scareware, Trojan, Trojan-Banker, Trojan-Dropper, Trojan-SMS, Trojan-Spy, and Zero-day. This dataset serves as a valuable resource for robust and scalable Android malware detection research. The CIC-AndMal2020 dataset consists of both static and dynamic features extracted from APK files. In this context we only consider the static features for training our proposed model. The static features include permissions, API calls, intents, and strings, while the dynamic features include system calls, network traffic, and file operations. The static features are extracted from the APK files without executing them, providing insights into the permissions, intents, and strings used by the applications. The features were extracted using the AndroidAppLyzer\footnote{https://github.com/ahlashkari/AndroidAppLyzer} tool adopted in \citep{rahali2020didroid} and \citep{keyes2021entroplyzer}. Using the AndroidAppLyzer, the static features were systematically extracted from Android application packages (.apk) through a structured three-step process: feature collection, feature capturing, and feature extraction and combination. The approach involved reverse engineering .apk files using apktool\footnote{https://github.com/iBotPeaches/Apktool} to extract critical components from AndroidManifest.xml, including activities, permissions, metadata, and system features. A custom script was employed to handle Android's proprietary binary XML format and capture feature data as text strings. The extracted features were then transformed into numerical vectors, represented either as binary indicators (1 for presence, 0 for absence) or frequency-based counts. This method ensures a robust representation of app behavior, facilitating effective training for the malware detection model.

\subsubsection{BODMAS Dataset}
\par The BODMAS dataset \citep{yand2021bodmas}, short for Blue Hexagon Open Dataset for Malware Analysis, is a comprehensive dataset designed to facilitate research in machine learning-based malware detection and analysis. Created in collaboration with Blue Hexagon, this dataset addresses gaps in existing datasets by providing timestamped malware samples and well-curated family information, enabling researchers to study critical challenges such as concept drift and malware family evolution. The dataset contains 57,293 malware samples and 77,142 benign samples, totaling 134,435 samples, collected from August 2019 to September 2020. Malware samples are sourced monthly from a security company’s internal database, ensuring their relevance and recency. For each malware sample, the dataset provides its SHA-256 hash, the original PE binary, and a pre-extracted feature vector. For benign samples, only the SHA-256 hash and feature vector are included due to copyright restrictions. The feature vectors follow the same format as those in the EMBER, ensuring compatibility for researchers working across multiple datasets. A distinguishing feature of BODMAS is its curated malware family information, covering 581 malware families. This labeling is based on in-house scripts that analyze antivirus vendor verdicts and are continuously updated by Blue Hexagon’s threat team. The dataset encompasses 14 malware categories, including prevalent types such as Trojans (29,972 samples), Worms (16,697 samples), Backdoors (7,331 samples), Downloaders (1,031 samples), and Ransomware (821 samples). Features for each sample are extracted using the LIEF project\footnote{https://github.com/elastic/ember}, generating a 2381-dimensional feature vector compatible with ML/DL models. This pre-extracted format accelerates research workflows by enabling immediate experimentation on labeled data. Additionally, the inclusion of timestamps and curated family information allows for longitudinal studies on malware evolution and the impact of concept drift.

\subsubsection{EMBOD Dataset}
\par The EMBOD dataset, newly introduced in this research, is a custom dataset designed to advance malware detection by combining the strengths of two widely recognized datasets: EMBER \citep{anderson2018ember} and BODMAS. By removing unlabeled samples from the EMBER dataset and integrating the remaining labeled samples with the BODMAS dataset, EMBOD provides a comprehensive and balanced dataset suitable for both static and dynamic malware analysis. The combined dataset contains a total of 934,311 samples, offering a rich foundation for machine learning (ML) and deep learning (DL) experiments. The EMBER 2018 dataset, originally comprising features from 1 million PE files scanned in or before 2018, included 300,000 each of malicious, benign, and unlabeled samples for training, along with 200,000 labeled test samples. For this work, only the labeled samples were retained, ensuring the integrity and relevance of the data for supervised learning tasks. Each sample is represented as a 2381-dimensional feature vector extracted using the LIEF project (version 0.9.0). These features include both parsed features, such as headers, imports, and sections, and format-agnostic features, such as histograms, entropy statistics, and strings, providing comprehensive static information about each sample.

\begin{figure}[!ht]
\centering
\includegraphics[scale=0.35]{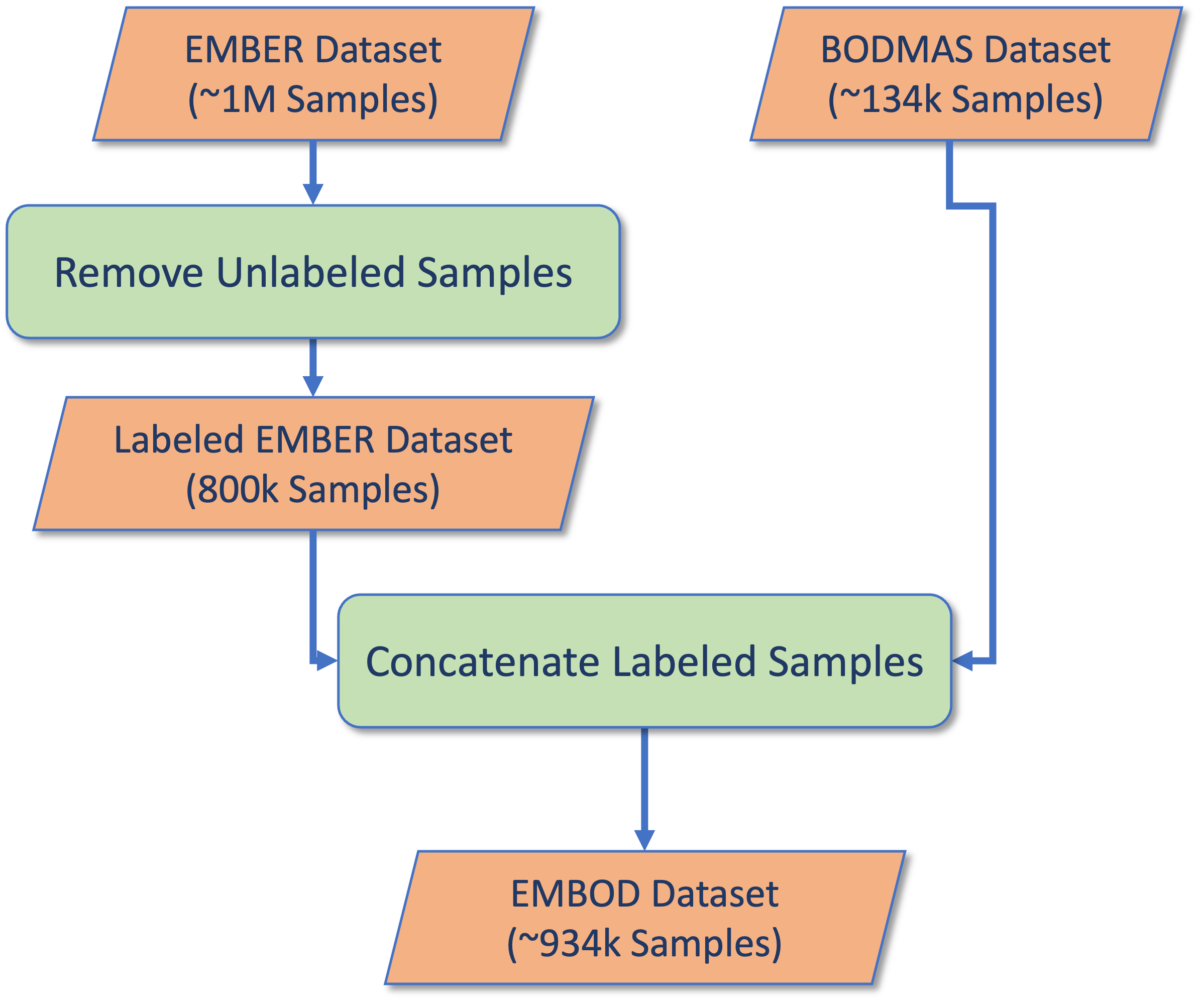} 
\caption{EMBOD Dataset Creation Process\label{fig:embod-dataset-creation}}
\end{figure}

By merging two benchmark datasets, EMBOD provides the following advantages:
\begin{itemize}
\item \emph{Balanced and Large-Scale Dataset}: With 934,311 labeled samples, EMBOD offers extensive coverage of malware and benign files, supporting robust model training and evaluation.
\item \emph{Temporal and Contextual Diversity}: Incorporating malware samples from 2018 and earlier (via EMBER) and recent samples from 2019–2020 (via BODMAS) enables the study of concept drift and evolving malware characteristics.
\item \emph{Curated Family Information}: The inclusion of malware family labels from BODMAS enriches the dataset, allowing for granular analysis of family-based patterns and behaviors.
\item \emph{Unified Feature Representation}: Both datasets utilize the same feature extraction methodology, ensuring consistency and compatibility for ML/DL frameworks.
\end{itemize}
The EMBOD dataset, introduced in this paper, fills critical gaps in existing malware datasets by combining historical breadth with modern relevance. Its size, diversity, and curated nature make it a powerful resource for developing and benchmarking state-of-the-art malware detection frameworks. The dataset has been made available for research community to facilitate the development of robust and scalable malware detection models \footnote{https://www.kaggle.com/datasets/ajvadhaneef/embod-all/}. The process of merging the EMBER and BODMAS datasets to create the EMBOD dataset is illustrated in Figure \ref{fig:embod-dataset-creation}.

\subsection{Data Preprocessing} 
The preprocessing phase consists of a series of steps aimed at data cleaning, transformation, and standardization to facilitate effective machine learning model training.
\subsubsection{Preprocessing of CIC-AndMal dataset}
\begin{figure}[!ht]
\centering
\includegraphics[scale=0.35]{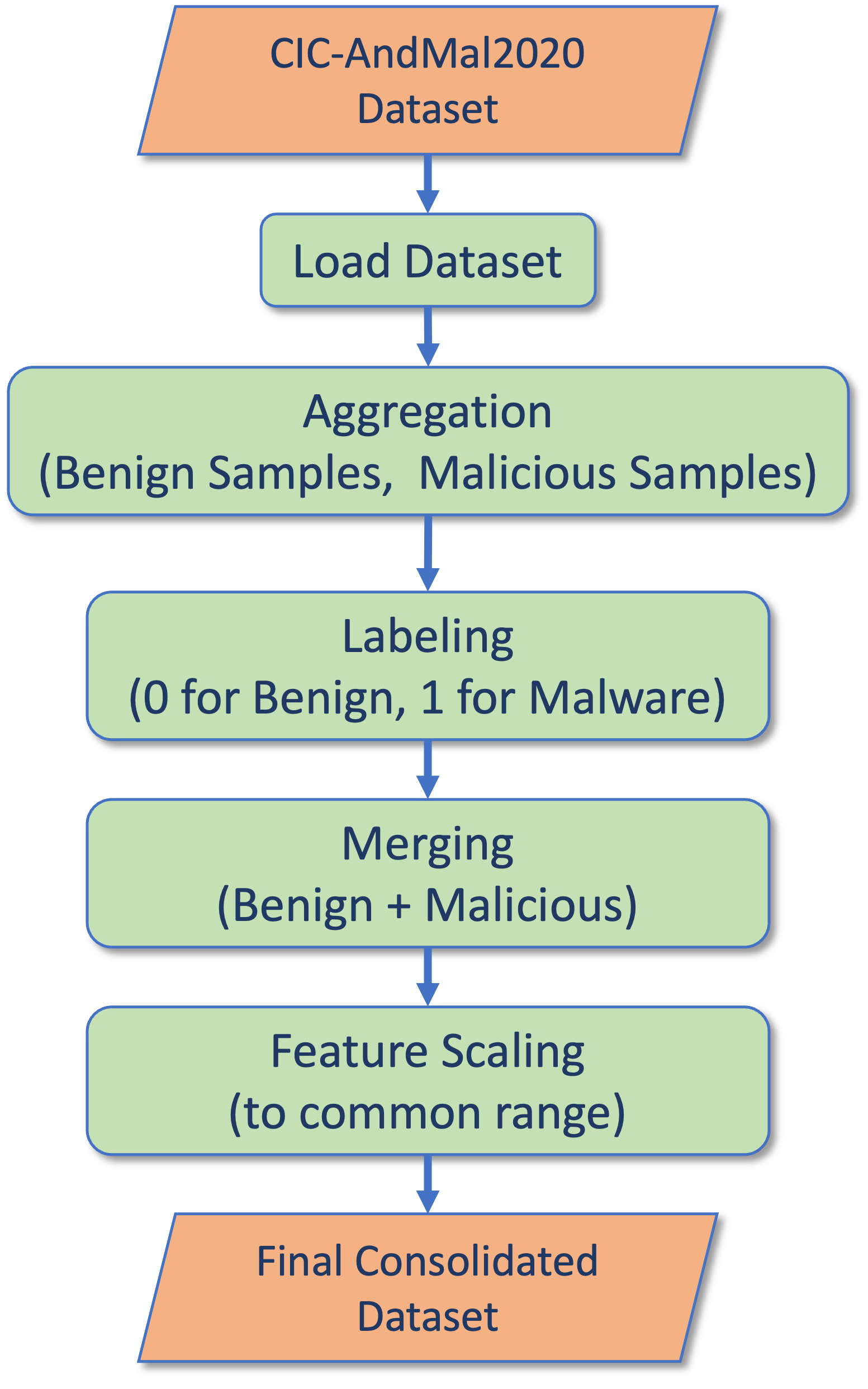} 
\caption{Data Preprocessing Pipeline - CIC-AndMal2020 Dataset\label{fig:andmal-2020-preprocessing}}
\end{figure}

The CIC-AndMal2020 dataset contains static features organized into separate directories for benign and malicious samples, with the malicious samples further categorized into different malware families. To streamline the preprocessing pipeline and enable binary classification, all benign and malicious samples were merged into a single dataset, with labels assigned as 0 for benign and 1 for malicious. This consolidation step ensures a consistent and uniform structure, facilitating efficient downstream processing, feature extraction, and model training. Additionally, feature scaling was applied to normalize all feature values to a common range, enhancing model generalization. The overall data preprocessing pipeline for the CIC-AndMal2020 dataset is illustrated in Figure \ref{fig:andmal-2020-preprocessing}.

\subsubsection{Preprocessing of BODMAS and EMBOD datasets}
For the BODMAS and EMBOD datasets, preprocessing primarily involved feature scaling to ensure all numerical values were within a standardized range. This transformation is crucial for maintaining uniformity across different datasets and improving the model’s ability to learn meaningful patterns. The scaled features were subsequently used for model training and evaluation, ensuring a consistent input format across all experiments.

\subsection{Evaluation Metrics}
\label{sec:evaluation_metrics}
We evaluate our proposed method in terms of Accuracy($A_c$), Precision($P_c$), Recall($R_c$), F1 score($F1_c$), Matthews Correlation Coefficient($M_c$), and Area Under the Receiver Operating Characteristic Curve ($AUC-ROC$). These metrics are calculated based on the number of True Positives (TP), True Negatives (TN), False Positives (FP), and False Negatives (FN). These metrics provide insight into various aspects of the model's ability to correctly classify instances from our dataset. The chosen evaluation metrics reflect the multifaceted performance requirements of a malware detection system. 

\noindent\textbf{Accuracy($\mathcal{A}_c$):} Accuracy measures the proportion of correct predictions (both true positives and true negatives) out of the total number of predictions made. The accuracy is defined as:
\begin{equation}
    \mathcal{A}_c = \frac{TP + TN}{TP + TN + FP + FN}
\end{equation}
\textbf{Precision ($\mathcal{P}_c$):} Precision reflects the ability of the model to correctly identify positive cases out of all the cases it predicted as positive. In other words, it measures the proportion of true positive predictions among all positive predictions made by the model.  The precision is defined as:
\begin{equation}
    \mathcal{P}_c = \frac{TP}{TP + FP}
\end{equation}
\textbf{Recall ($\mathcal{R}_c$):} Recall measures the proportion of true positive predictions out of all actual positive instances in the dataset. It focuses on minimizing false negatives. 
TPR measures the proportion of actual positive cases that were correctly identified as positive by the model. The recall  is defined as:

\begin{equation}
    \mathcal{R}_c = \frac{TP}{TP + FN}
\end{equation}  
\textbf{F1-Score($\mathcal{F}1_c$):} The F1-Score is the harmonic mean of precision and recall. It provides a balance between precision and recall, especially useful when dealing with imbalanced datasets where one class is much less prevalent than the other. The F1-score is defined as:
\begin{equation}
    \mathcal{F}1_c = 2 * \frac{\mathcal{P}_c * \mathcal{R}_c}{\mathcal{P}_c + \mathcal{R}_c}
\end{equation}
\textbf{Matthews Correlation Coefficient ($\mathcal{M}_c$):} The Matthews Correlation Coefficient (MCC) is a balanced metric that measures the quality of binary classifications, taking into account true positives (TP), true negatives (TN), false positives (FP), and false negatives (FN). Unlike accuracy, which can be misleading in imbalanced datasets, MCC provides a more informative and fair evaluation of classifier performance. An MCC value close to +1 indicates perfect classification, 0 suggests random predictions, and -1 represents completely incorrect classifications. The MCC is defined as: 
\begin{equation}
    \mathcal{M}_c = \frac{(TP \cdot TN) - (FP \cdot FN)}{\sqrt{(TP + FP)(TP + FN)(TN + FP)(TN + FN)}}
\end{equation}
\textbf{Area Under the Receiver Operating Characteristic Curve ($AUC-ROC$):} The AUC-ROC metric evaluates the model’s ability to distinguish between malware and benign samples across varying classification thresholds. AUC-ROC is derived from the ROC curve, which plots the True Positive Rate (TPR) against the False Positive Rate (FPR), which are defined as: 
\begin{equation}
TPR = \frac{TP}{TP + FN}
\end{equation}
\begin{equation}
FPR = \frac{FP}{FP + TN}
\end{equation}
The AUC score is the integral of the ROC curve, represeted as:
\begin{equation}
    AUC-ROC = \int_{0}^{1} TPR(FPR) dFPR
\end{equation}

AUC score quantifies the model’s discrimination power, where a score of 1.0 indicates a perfect classifier and 0.5 suggests random guessing. Unlike fixed-threshold metrics, AUC-ROC is threshold-independent, making it highly suitable for evaluating the trade-off between false positives and false negatives, a critical aspect of malware detection.
\subsection{Implementation Details}
The experiments were conducted using Python on an Nvidia DGX Station A100, which features an AMD EPYC 7742 64-core processor, 4× Tesla A100 GPUs with 40GB of memory each, and 512GB of DDR4 RAM. This high-performance setup ensures efficient processing of large-scale datasets and enables the computationally intensive training of deep learning models. The proposed MAML-based model was implemented using Python, along with TensorFlow 2.0 and Keras, both of which provide a comprehensive framework for deep learning and meta-learning. The training process involved optimizing model parameters using the Adam optimizer, with the key hyperparameters: inner-loop learning rate ($\alpha = 0.0001$) and outer-loop learning rate ($\beta = 0.001$). The model was trained using the binary cross-entropy (BCE) loss function, which is well-suited for binary classification tasks. The number of outer-loop iterations, number of samples per task, and the sizes of the support and query sets varied depending on the dataset used. The parameter settings are summarized in Table \ref{tab:hyperparameter_settings}.

\begin{table*}[!ht]
    \centering
    \begin{threeparttable}
    \caption{Hyperparameter settings}
    \label{tab:hyperparameter_settings}
    \begin{tabular}{lccc} 
        \toprule
        \textbf{Parameter} & \textbf{CIC-AndMal}  &  \textbf{BODMAS}  & \textbf{EMBOD} \\
        \midrule
        Number of Chunks ($k$) & 9 & 24	& 17	\\
        Chunk Size ($p$) & 0.20 & 0.05	& 0.06	\\
        Overlap between Chunks ($q$) & 0.20 & 0.10	& 0.10	\\
        Threshold for top 100 features ($\tau$) & 0.00014 & 0.00716	& 0.01638	\\
        Inner-loop Learning Rate ($\beta$) & 0.0001 & 0.0001	&0.0001	\\
        Outer-loop Learning Rate ($\beta$) & 0.001 & 0.001	&0.001	\\
        Number of Iterations (Outer Loop) & 1000 & 500 & 1000 \\
        Number of Samples per Task ($\left | \mathcal{D}_{task} \right |$)& 10000 & 5000 & 30000 \\
        Support Set Size  ($\left | \mathcal{P}_{i} \right |$) & 5000 & 2500 & 15000 \\
        Query Set Size  ($\left | \mathcal{Q}_{i} \right |$) & 5000 & 2500 & 15000 \\
        Loss Function & BCE & BCE & BCE \\
        Optimizer & Adam & Adam & Adam \\
        \bottomrule
    \end{tabular}
    \begin{tablenotes}
        \footnotesize
        \item BCE: Binary Cross-Entropy
      \end{tablenotes}
    \end{threeparttable}
\end{table*}

\subsection{Comparison with State-of-the-art Approaches}
The performance comparison of the proposed method on CIC-AndMal2020 and BODMAS datasets, with existing state-of-the-art approaches is presented in Table \ref{tab:model_comparison_andmal} and Table \ref{tab:model_comparison_bodmas}, respectively. The methods were evaluated in terms of key performance metrics including accuracy, precision, recall, F1-score, MCC and AUC mentioned in Section \ref{sec:evaluation_metrics}. The absence of metrics in the tables is marked as "-", indicates that the respective studies did not report those metrics.

\begin{table*}[!ht]
    \centering
    \begin{threeparttable}
    \caption{Comparison of performance of state-of-the-art models on CIC-AndMal2020 dataset}
    \label{tab:model_comparison_andmal}
    \begin{tabular}{p{3.5cm}ccccccccc} 
        \toprule
        \textbf{Reference} & \textbf{Features}  &  \textbf{Classification}  & \textbf{Method}&\textbf{$\mathcal{A}_c$} & \textbf{$\mathcal{P}_c$} & \textbf{$\mathcal{R}_c$} &\textbf{$\mathcal{F}_c$}   &\textbf{$\mathcal{M}_c$}  &\textbf{AUC} \\
        \midrule
        \cite{rahali2020didroid}  & Static  & M  & CNN &93.36 & 93.4 & 93.4 & 93.4 & - & - \\
        \cite{batouche2021comprehensive}  & Static  & M  & RF & 89 & 90 & 90 & 90  & - & -\\
        \cite{ullah2022trojandetector}  & Hybrid & B & SVM & 96.64 &98 & 96.7  & - & - & -\\
        \cite{al2022parallel}  & Static  & B & PDL-FEMC & 97.6 &97.7 & - & - & - & - \\
        \cite{musikawan2022enhanced}  & Static  &M & Ensemble ML + DNN &78.82 &79.13 & 78.82 & 78.63  & - & -\\
        \cite{musikawan2022enhanced}  & Static  & B & Ensemble ML + DNN & 97.72 & 97.73 & 97.72 & 97.72  & - & -\\
        \cite{islam2023android} & Static  & M  & Ensemble ML &95 & - & - & - & - & -\\
        \cite{chopra2023energy}  & Static  & B & Transfer Learning & 97.19 &- & - & - & - & -\\
        \cite{singh2024s} & Static  & B & Attention + MLP & 89.82 & - & - & - & - & -\\
        \rowcolor{grey1} \textbf{MeLeMaD (proposed)}    & \textbf{Static} & \textbf{B} & \textbf{CFSGB+MAML} & \textbf{98.04} & \textbf{97.59} & \textbf{98.52} & \textbf{98.05}  & \textbf{0.9609} & \textbf{0.9959} \\
        \bottomrule
    \end{tabular}
    \begin{tablenotes}
        \footnotesize
        \item B: Binary Classification; M: Multiclass (Category) Classification.
      \end{tablenotes}
    \end{threeparttable}
\end{table*}

Table \ref{tab:model_comparison_andmal} presents the perforamnce comparison of the MeLeMaD framework with existing approaches on the CIC-AndMal2020 dataset. The MeLeMaD framework achieves the highest classification accuracy (98.04\%) compared to existing methods, demonstrating its effectiveness in detecting malware and benign samples. It also outperforms other models in recall (98.52\%), highlighting its superior ability to correctly identify malware without missing true positives. The F1-score (98.05\%) and precision (97.59\%) indicate a well-balanced detection capability, minimizing both false positives and false negatives. The Matthews correlation coefficient (0.9609) suggests that the model makes highly reliable predictions across both malware and benign classes. Additionally, the AUC (0.9959) reflects the framework's excellent discriminative power, showing that it can effectively differentiate between malicious and benign samples. Compared to prior approaches, our model significantly improves upon static feature-based classifiers. This suggests that our feature selection technique, Chunk-wise Feature Selection based on Gradient Boosting (CFSGB), effectively selects the most relevant static features, compensating for the lack of dynamic features while maintaining superior performance. The results demonstrate the robustness and adaptability of the proposed MeLeMaD framework in detecting Android malware, even with limited static features.

\begin{table*}[!ht]
    \centering
    \begin{threeparttable}
    \caption{Comparison of performance of state-of-the-art models on BODMAS dataset}
    \label{tab:model_comparison_bodmas}
    \begin{tabular}{p{3.5cm}ccp{3cm}cccccccc} 
        \toprule
        \textbf{Reference}  & \textbf{Features} &  \textbf{Classification}  & \textbf{Method} &\textbf{$\mathcal{A}_c$} & \textbf{$\mathcal{P}_c$} & \textbf{$\mathcal{R}_c$} &\textbf{$\mathcal{F}1_c$}   &\textbf{$\mathcal{M}_c$}  &\textbf{AUC}\\
        \midrule
        \cite{ramadhan2021analysis}  & Static & B  & Hybrid ML (LGBM + XGB + LR) & 99.62 & 99.61 & 99.50 & 99.56 & - & -\\
        \cite{lu2022self}   & Static  & M & SeqConvAttn & 96.89 &- & - & 96.9 &- & - \\
        \cite{lu2022self}   & Static  & B & ImgConvAttn & 97 &- & - & 96.99 & - & - \\
        \cite{wang2022measurement}    & Hybrid & M & RF & 83.5 & 80.8 & - & 81.4 & - & - \\
        \cite{al2022parallel} & Static  & B  & PDL-FEMC &97.6 & 97.7 & 98 & 96.7 & - & -\\
        \cite{hussain2022hierarchical}   & Static   & B & RF & 99.48 & 99.52 & 99.42 & 93.57  \\
        \cite{hao2022eii}  & Static  & M & CNN+SPP & 99.4 &- & - & 0.91 & - & -  \\
        \cite{hai2023proposed}   & Static  & B & CNN (Inception V3) & 96.27 & 99.3 & 96.56 & 97.91 & - & -  \\
        \cite{manikandaraja2023rapidrift}   & Static & B & LightGBM & 98 & - & - & 96.1 & - & -  \\
        \cite{dener2023clustering}    & Static & B  & Clustering + Ensemble ML & 99.75 & - & - & 99.77 & - & -\\
        \cite{bhardwaj2024overcoming}   & Static  & B & MD-ADA & 99.29 & - & - & 99.13 & - & - \\
        \cite{robinette2024case}    & Static  & B & DNN & 99 & 99 & 99 & 99  & - & -\\
        \cite{buriro2024malware}   & Static  & B & RF & 99.73 &- & - & -  & - & -\\
        \cite{xiong2024modified}   & Static  & B & QCNN & 95.60 & 91.99 & 99.90 & 95.78  & - & -\\
        \cite{farfoura2025novel}   & Static  & B & LightGBM & 99 & 99 & 99 & 99  & - & -\\
        \rowcolor{grey1} \textbf{MeLeMaD (proposed)}    & \textbf{Static} & \textbf{B} & \textbf{CFSGB+MAML} & \textbf{99.97} & \textbf{99.99} & \textbf{99.94} & \textbf{99.97}  & \textbf{0.9994} & \textbf{1.000} \\
        \bottomrule
    \end{tabular}
    \begin{tablenotes}
        \footnotesize
        \item B: Binary Classification; M: Multiclass (Category) Classification.
      \end{tablenotes}
    \end{threeparttable}
\end{table*}

Table
\ref{tab:model_comparison_bodmas} showcases the performance comparison of the MeLeMaD framework with existing approaches on the BODMAS dataset. The MeLeMaD achieves the highest accuracy of 99.97\%, outperforming other methods in terms of precision (99.99\%), recall (99.94\%), F1-score (99.97\%), and Matthews correlation coefficient (0.9994). The AUC score of 1.000 indicates the model's exceptional discriminative power, highlighting its ability to distinguish between malware and benign samples effectively. The results demonstrate the effectiveness of the proposed approach in enhancing malware detection capabilities.

The proposed MeLeMaD framework demonstrates state-of-the-art performance across both the CIC-AndMal2020 (Android malware) and BODMAS (Windows malware) datasets, highlighting its adaptability and effectiveness in malware detection. These results establish the generalization ability of our framework across different malware ecosystems—Android (AndMal2020) and Windows (BODMAS)—demonstrating that optimized static feature selection (CFSGB) and meta-learning-based adaptability (MAML) significantly enhance detection accuracy and robustness across diverse environments.

\begin{table*}[!ht]
    \centering
    \begin{threeparttable}
    \caption{Comparison of performance of selected ML models on EMBOD dataset}
    \label{tab:model_comparison_embod}
    \begin{tabular}{lp{5.5cm}cccccc} 
        \toprule
        
          \textbf{Classifier Model} &\textbf{Hyperparameters}  &\textbf{$\mathcal{A}_c$} & \textbf{$\mathcal{P}_c$} & \textbf{$\mathcal{R}_c$} &\textbf{$\mathcal{F}1_c$}   &\textbf{$\mathcal{M}_c$}  &\textbf{AUC}\\
        \midrule
        
        Logistic Regression &\small{\emph{\{``solver":``lbfgs", ``penalty":``l2"\}}} &82.14 & 81.44 & 82.24 & 81.84 & 0.6427 & 0.9056 \\ \hline
        Random Forest &\small{\emph{\{``criterion":``gini", ``n\_estimators":100\}}} &96.53 & 99.70 & 95.15 & 96.41 & 0.9309 & 0.9946 \\ \hline
        SGD Classifier & \small{\emph{\{``loss":``modified\_huber",``learning\_rate": ``constant"\}}} &80.31 & 80.31 & 80.31 & 80.31 & 0.6067 & 0.8925 \\ \hline
        SVM & \small{\emph{\{``kernel":``rbf", ``gamma":``scale"\}}}&92.65 & 93.04 & 91.83 & 92.44 & 0.8529 & 0.9756 \\ \hline
        LightGBM &\small{\emph{\{``n\_estimators": 200,``learning\_rate":0.1\}}} &94.97 & 94.97 & 94.97 & 94.97 & 0.8994 & 0.9898 \\ \hline
        GradientBoosting &\small{\emph{\{``n\_estimators": 100, ``learning\_rate": 0.1\}}} &87.63 & 86.94 & 87.94 & 87.44 & 0.7526 & 0.9513 \\ \hline
        XGBoost & \small{\emph{\{``n\_estimators":100, ``booster":``gbtree"\}}}&94.16 & 94.49 & 93.52 & 94.00 & 0.8832 & 0.9870 \\ \hline
        AdaBoost & \small{\emph{\{``n\_estimators":50,``learning\_rate":1.0\}}}&83.32 & 82.02 & 84.40 & 83.30 & 0.6666 & 0.9142 \\ \hline
        ExtraTrees & \small{\emph{\{``n\_estimators":100, ``criterion":``gini"\}}}&96.70 & 97.98 & 95.21 & 96.58 & 0.9343 & 0.9952 \\ \hline
        Bagging Classifier & \small{\emph{\{``n\_estimators":10, ``bootstrap":``True" \}}} &95.32 & 96.69 & 93.64 & 95.14 & 0.9068 & 0.9891 \\ \hline
        Voting Ensemble (KNN, RF) & \small{\emph{KNN\{``metric":``euclidean",``n\_neighbors": 82\}; RF\{``criterion":``gini",``n\_estimators": 200\}}} &93.44 & 93.44 & 93.44 & 93.44 & 0.8690 & 0.9844 \\  \hline
        Voting Ensemble (RF, SVM,XGB)&  \small{\emph{RF\{``criterion":``gini",``n\_estimators":100\}; SVM\{``kernel":``rbf", ``gamma":``scale"\}; XGB\{``n\_estimators":100\}}} &95.24 & 95.64 & 94.58 & 95.11 & 0.9047 & 0.9918 \\ \hline
        \rowcolor{grey1} \textbf{MeLeMaD (proposed)} & As indicated in Table \ref{tab:hyperparameter_settings} & \textbf{97.85} & \textbf{97.55} & \textbf{98.06} & \textbf{97.81}  & \textbf{0.9570} & \textbf{0.9973} \\
        \bottomrule \hline
    \end{tabular}
    \begin{tablenotes}
        \footnotesize
        \item RF: Random Forest Classifier; KNN: K-Nearest Neighour Classifier; SVM: Support Vector Machine; SGD: Stochastic Gradient Descent; 
      \end{tablenotes}
    \end{threeparttable}
\end{table*}

\subsection{Comparison of EMBOD dataset with selected ML models}
To further evaluate the robustness and generalization capability of the MeLeMaD framework, we conducted experiments on the newly introduced EMBOD dataset, which integrates the EMBER and BODMAS datasets. The results of various machine learning classifiers, including tree-based models, ensemble learning, and the MeLeMaD framework, are presented in Table \ref{tab:model_comparison_embod}. Given the high-dimensional nature of the dataset, Principal Component Analysis (PCA) with 100 components was employed for dimensionality reduction in all baseline models. In contrast, the proposed MeLeMaD framework utilized the newly introduced Chunk-wise Feature Selection based on Gradient Boosting (CFSGB) technique for feature selection, enhancing its optimized approach. As shown in Table \ref{tab:model_comparison_embod}, ensemble-based methods, particularly Random Forest, ExtraTrees, Bagging Classifier, and Voting Ensembles—outperform traditional classifiers such as Logistic Regression and Support Vector Machines. The Random Forest classifier achieved an accuracy of 96.53\%, an MCC of 0.9309, and an AUC of 0.9946, demonstrating its robustness. Similarly, the ExtraTrees classifier attained the highest accuracy among baseline models (96.70\%), confirming its effectiveness in malware classification. The MeLeMaD framework achieves the highest classification performance, with an accuracy of 97.85\%, surpassing all baseline models. Additionally, it exhibits the highest recall (98.06\%), ensuring minimal false negatives—an essential factor in cybersecurity applications. Furthermore, its AUC score of 0.9973 underscores its superior discriminative capability between benign and malicious samples, solidifying its effectiveness as a state-of-the-art malware detection framework.
  
\begin{figure*}[!ht]
\centering
\includegraphics[width=1\textwidth]{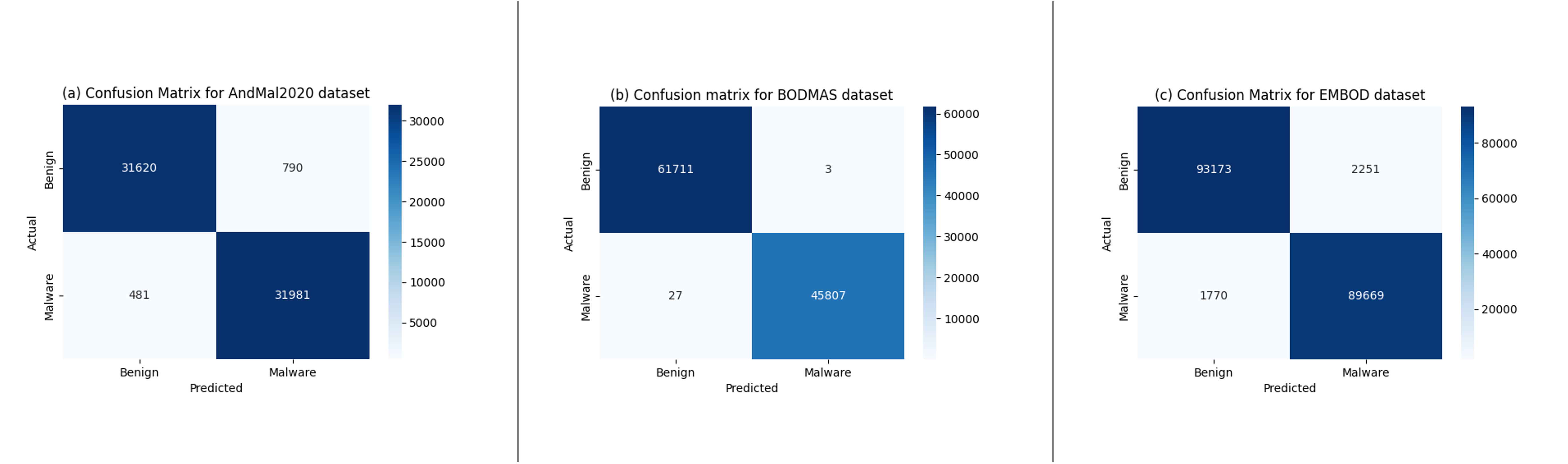} 
\caption{Confusion matrices \label{fig:cm-all} illustrating the performance of MeLeMaD framework on (a) AndMal2020, (b) BODMAS, and (c) EMBOD datasets.}
\end{figure*}

\begin{figure*}[!ht]
\centering
\includegraphics[width=1\textwidth]{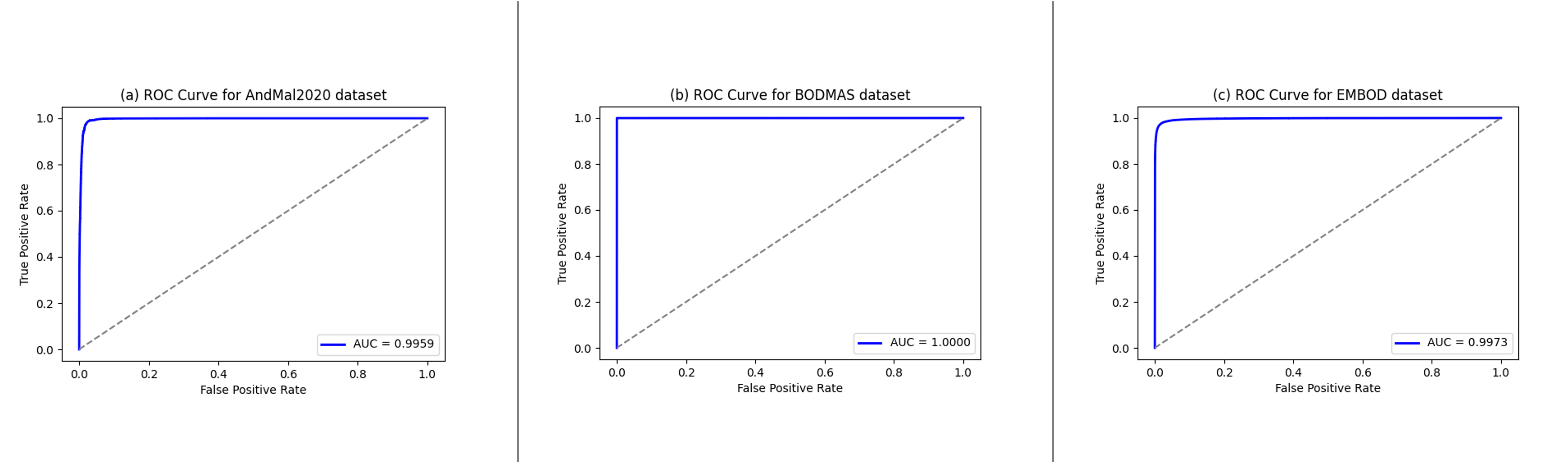} 
\caption{ROC curves showing the classification performance of the MeLeMaD framework on (a) AndMal2020, (b) BODMAS, and (c) EMBOD datasets.\label{fig:roc-all}}.
\end{figure*}

To provide a comprehensive visualization of the performance of MeLeMaD framework across different datasets, we include confusion matrices (Figure \ref{fig:cm-all}) and AUC–ROC curves (Figure \ref{fig:roc-all}) for AndMal2020, BODMAS, and EMBOD. The confusion matrices reveal the distribution of true positives, false positives, true negatives, and false negatives, illustrating how effectively the model distinguishes benign and malicious samples. A high volume of true positives alongside minimal false negatives indicates that the approach effectively captures malicious behavior, while a low number of false positives suggests strong specificity against benign samples. The AUC–ROC curves offer an additional perspective on classification quality by plotting the true positive rate against the false positive rate, summarizing the model’s overall ability to discriminate between classes. Notably, an AUC approaching 1.0 underscores the model’s high discriminative power, while any visible differences in curve shapes across datasets provide insights into how generalizable the model is across diverse malware samples. By examining both the confusion matrices and AUC–ROC curves together, we gain a nuanced understanding of the framework’s precision, recall, and overall robustness across multiple threat landscapes.

\subsection{Discussion}
The experimental results presented in the previous sections demonstrate the effectiveness of the proposed MeLeMaD framework in detecting malware across different datasets. The framework consistently outperforms existing state-of-the-art approaches on both the CIC-AndMal2020 and BODMAS datasets, achieving the highest accuracy, precision, recall, F1-score, MCC, and AUC. This superior performance can be attributed to the synergistic integration of the Chunk-wise Feature Selection based on Gradient Boosting (CFSGB) technique and the Model-Agnostic Meta-Learning (MAML) approach. The CFSGB technique effectively selects the most relevant features from the static feature set, particularly in the context of large, high-dimensional datasets including CIC-AndMal2020 and EMBOD. The MAML approach further improves the model's adaptability, allowing it to quickly learn from new tasks with limited data. The results on the EMBOD dataset further validate the robustness and generalization capability of the MeLeMaD framework. The framework achieves the highest classification performance compared to state-of-the-art methods.  The confusion matrices and AUC-ROC curves provide additional insights into the model's performance, highlighting its ability to accurately classify benign and malicious samples while minimizing false positives and false negatives. 

The MeLeMaD framework's adaptability is primarily driven by the MAML approach, which enables the model to learn a good initialization that can be fine-tuned with minimal data from new tasks. In practical terms, when a new malware variant is encountered, the framework can leverage its pre-trained knowledge and quickly adapt to the new threat with only a few labeled examples. This few-shot learning capability is particularly valuable in real-world scenarios where obtaining large amounts of labeled data for new malware types can be challenging. For instance, in an enterprise environment, when a novel malware strain is detected, security analysts can provide a small set of labeled samples to the MeLeMaD framework, which can then rapidly adapt and enhance its detection capabilities for that specific threat. This adaptability not only improves the efficiency of malware detection but also reduces the time and resources required for retraining traditional models from scratch. The ability to generalize across different datasets and malware types further enhances the framework's utility in diverse cybersecurity contexts. As cyber threats continue to evolve, the MeLeMaD framework's combination of optimized feature selection and meta-learning positions it as a promising solution for proactive and adaptive malware detection. The ability of the MeLeMaD framework to generalize across different datasets and malware types underscores its potential as a robust solution for real-world malware detection challenges. The framework's strengths lie in its adaptability, efficiency, and high performance across various metrics. However, it is essential to acknowledge potential limitations, such as the reliance on the quality of the initial training data and the computational complexity associated with meta-learning approaches. Future research could explore further optimizations in feature selection techniques and investigate the integration of additional data sources, such as dynamic analysis features, to enhance the framework's robustness against sophisticated malware variants. Overall, the experimental results highlight the effectiveness of the MeLeMaD framework in enhancing malware detection capabilities, making it a promising approach for future research and practical applications in cybersecurity.

 \subsection{Ablation Studies}
 A systematic ablation experiments have been conducted to analyse the effectivenes of CFSGB and MAML classification on CIC-AndMal2020 and EMBOD datasets. The objective of these experiments is to quantify the impact of each component of MeLeMaD and validate the significance of  their integration. The ablation exeperimants were performed  using the following configurations:
\begin{enumerate}
\item \textbf{CFSGB + Conventional Classifiers}: This configuration employs the Chunk-wise Feature Selection based on Gradient Boosting (CFSGB) technique for feature selection, followed by a conventional classifier (e.g., Randfom Forest, LightGBM etc.) for classification.
\item \textbf{MAML + Baseline Feature Selection}: This configuration utilizes Model-Agnostic Meta-Learning (MAML) for classification, but without the CFSGB feature selection. Instead, it uses a baseline feature selection method (e.g., Variance Threshold, ANOVA etc.).
\end{enumerate}
 \subsubsection{Effectiveness of CFSGB}

 The effectiveness of the Chunk-wise Feature Selection based on Gradient Boosting (CFSGB) technique is evaluated by comparing its performance with conventional classifiers on the CIC-AndMal2020 and EMBOD datasets. The results are presented in Table \ref{tab:ablation_cfsgb}. The results indicate that the CFSGB technique significantly enhances the performance of conventional classifiers. For instance, on the CIC-AndMal2020 dataset, the Random Forest classifier achieves an accuracy of 96.53\% with CFSGB, compared to lower accuracies when using traditional feature selection methods. Similarly, on the EMBOD dataset, classifiers such as LightGBM and XGBoost show marked improvements in accuracy, precision, recall, F1-score, MCC, and AUC when paired with CFSGB. The results demonstrate that CFSGB effectively selects the most relevant features from the static feature set, leading to improved classification performance across various metrics. This highlights the importance of optimized feature selection in enhancing the capabilities of conventional classifiers in malware detection tasks.

\begin{table*}[!ht]
    \centering
        \small
    \begin{threeparttable}
    \caption{Ablation study results for CFSGB on EMBOD and CIC-AndMal2020 datasets}
    \label{tab:ablation_cfsgb}
    \renewcommand{\arraystretch}{1.2}
    \begin{tabular}{l|cccccc|cccccc}
        \toprule
        \multirow{2}{*}{\textbf{Configuration}} &
        \multicolumn{6}{c|}{\textbf{EMBOD}} & 
        \multicolumn{6}{c}{\textbf{AndMal2020}} \\
        \cmidrule{2-13}
         & $\mathcal{A}_c$ & $\mathcal{P}_c$ & $\mathcal{R}_c$ & $\mathcal{F}_c$ & $\mathcal{M}_c$  & \emph{AUC} & $\mathcal{A}_c$ & $\mathcal{P}_c$ & $\mathcal{R}_c$ & $\mathcal{F}_c$ & $\mathcal{M}_c$ & \emph{AUC} \\
        \midrule
        CFSGB + LR & 81.16 & 79.20 & 83.41 & 81.25 & 0.6243 & 0.9014 & 95.51 & 94.30 & 96.87 & 95.57 & 0.9105 & 0.9889 \\ \midrule
        CFSGB + RF & 97.04 & 96.79 & 97.17 & 96.98 & 0.9408 & 0.9704 & 97.15 & 96.66 & 97.67 & 97.16 & 0.9430 & 0.9715 \\ \midrule
        CFSGB + KNN & 96.20 & 96.03 & 96.22 & 96.12 & 0.9240 & 0.9872 & 95.31 & 93.97 & 96.84 & 95.39 & 0.9067 & 0.9798 \\ \midrule
        CFSGB + AdaBoost & 90.43 & 89.40 & 91.26 & 90.32 & 0.8087 & 0.9678 & 95.84 & 94.81 & 96.99 & 95.89 & 0.9171 & 0.9919 \\ \midrule
        CFSGB + GB & 91.23 & 90.53 & 91.67 & 91.09 & 0.8246 & 0.9738 & 96.24 & 95.35 & 97.24 & 96.28 & 0.9251 & 0.9922 \\ \midrule
        CFSGB + XGBoost & 96.59 & 96.70 & 96.31 & 96.50 & 0.9317 & 0.9947 & 97.79 & 97.24 & 98.37 & 97.80 & 0.9558 & 0.9949 \\ \midrule
        CFSGB + LightGBM & 95.04 & 95.01 & 94.84 & 94.92 & 0.9007 & 0.9900 & 97.48 & 96.91 & 98.08 & 97.49 & 0.9496 & 0.9944 \\ \midrule
        CFSGB + MLP & 95.66 & 96.09 & 95.01 & 95.54 & 0.9132 & 0.9902 & 97.45 & 97.42 & 97.47 & 97.45 & 0.9489 & 0.9949 \\ \midrule
        CFSGB + CatBoost & 93.09 & 92.69 & 93.22 & 92.96 & 0.8617 & 0.9828 & 96.90 & 96.00 & 97.89 & 96.94 & 0.9383 & 0.9947 \\ \midrule
        CFSGB+ VE & 95.88 & 95.26 & 96.38 & 95.82 & 0.9177 & 0.9874 & 96.83 & 95.54 & 98.25 & 96.87 & 0.9369 & 0.9942 \\ \midrule
        \rowcolor{grey1} 
        MeLeMaD  & 
        97.85 & 97.55 & 98.06 & 97.81 & 0.9570 & 0.9973 &
        98.04 & 97.59 & 98.52 & 98.05 & 0.9609 & 0.9959 \\ \bottomrule
    \end{tabular}
    \begin{tablenotes}
        \footnotesize
        \item LR: Logistic Regression; RF: Random Forest Classifier; KNN: K-Nearest Neighour Classifier; GB: Gradient Boosting; XGBoost: Extreme Gradient Boosting; MLP: Multilyaer Perceptron; VE: Voting Ensemble (LR, KNN).
      \end{tablenotes}
    \end{threeparttable}
\end{table*}

\subsubsection{Effectiveness of MAML classification}

    The effectiveness of Model-Agnostic Meta-Learning (MAML) classification is evaluated by comparing its performance with baseline feature selection methods on the CIC-AndMal2020 and EMBOD datasets. The results are presented in Table \ref{tab:ablation_maml}. The table shows that MAML classification significantly enhances the performance of various baseline feature selection methods across all evaluation metrics. For instance, on the EMBOD dataset, the combination of Variance Threshold (VT) with MAML achieves an accuracy of 96.33\%, compared to lower accuracies when VT is used with conventional classifiers (as seen in Table \ref{tab:ablation_cfsgb}). Similar improvements are observed in precision, recall, F1-score, MCC, and AUC across other feature selection methods such as Chi-Squared (CHI) and Analysis of Variance (ANOVA). On the CIC-AndMal2020 dataset, MAML classification consistently outperforms baseline feature selection methods, achieving high accuracy and robustness. The results demonstrate that MAML effectively adapts to the underlying data distribution, leading to improved classification performance. The ablation study confirms that MAML is a crucial component of the MeLeMaD framework, contributing significantly to its overall effectiveness in malware detection.

\begin{table*}[!ht]
    \centering
    \small
    \begin{threeparttable}
    \caption{Ablation study results for MAML on EMBOD and CIC-AndMal2020 datasets}
    \label{tab:ablation_maml}
    \renewcommand{\arraystretch}{1.2}
    \begin{tabular}{l|cccccc|cccccc}
        \toprule
        \multirow{2}{*}{\textbf{Configuration}} &
        \multicolumn{6}{c|}{\textbf{EMBOD}} & 
        \multicolumn{6}{c}{\textbf{AndMal2020}} \\
        \cmidrule{2-13}
         & $\mathcal{A}_c$ & $\mathcal{P}_c$ & $\mathcal{R}_c$ & $\mathcal{F}_c$ & $\mathcal{M}_c$  & \emph{AUC} & $\mathcal{A}_c$ & $\mathcal{P}_c$ & $\mathcal{R}_c$ & $\mathcal{F}_c$ & $\mathcal{M}_c$ & \emph{AUC} \\
        \midrule
        VT + MAML & 96.33 & 94.98 & 97.65 & 96.30 & 0.9269 & 0.9947 & 98.03 & 97.57 & 98.50 & 98.04 & 0.9606 & 0.9802 \\ \midrule
        CHI + MAML & 96.83 & 96.23 & 97.33 & 96.78 & 0.9366 & 0.9954 & 96.60 & 94.96 & 98.44 & 96.68 & 0.9327 & 0.9904 \\ \midrule
        ANOVA + MAML & 97.10 & 96.85 & 97.24 & 97.05 & 0.9421 & 0.9957 & 97.62 & 97.36 & 97.89 & 97.62 & 0.9523 & 0.9942 \\ \midrule
        \rowcolor{grey1} 
        MeLeMaD  & 97.85 & 97.55 & 98.06 & 97.81 & 0.9570 & 0.9973 & 98.04 & 97.59 & 98.52 & 98.05 & 0.9609 & 0.9959 \\
        \bottomrule
    \end{tabular}
    \begin{tablenotes}
        \footnotesize
        \item VT: Variance Threshold; CHI: Chi-Squared;  ANOVA: Analysis of Variance. 
      \end{tablenotes}
    \end{threeparttable}
\end{table*}

These experimantal results conclusively demonstrate that both CFSGB and MAML based classification are independently contribute to performance, with their combination achieving the highest accuracy and robustness. While MAML enhances adaptability and generalization, CFSGB optimizes feature selection, leading to a synergistic effect that significantly improves malware detection capabilities. The ablation studies validate the importance of each component in the MeLeMaD framework, highlighting their individual and combined contributions to achieving state-of-the-art performance in malware detection across diverse datasets.

\section{Conclusion}\label{sec:conclusion}
MeLeMaD, is a novel framework for malware detection that leverages the adaptability and generalization capabilities of Model-Agnostic Meta-Learning (MAML). By integrating MAML, the framework rapidly adapts to evolving malware threats with minimal retraining, effectively addressing the dynamic nature of malware. MeLeMaD incorporates a novel feature selection technique, Chunk-wise Feature Selection based on Gradient Boosting (CFSGB), which identifies and prioritizes the most relevant features in large-scale high-dimensional datasets, ensuring computational efficiency and robustness. The proposed dual-layered architecture, combining MAML for adaptability and CFSGB for efficient feature selection, outperforms existing state-of-the-art methods by achieving accuracies of 98.04\% on the CIC-AndMal2020 dataset and 99.97\% on the BODMAS dataset. The MeLeMaD's performance is further validated through extensive experiments on the newly introduced EMBOD dataset, achieving an accuracy of 97.85\%. The proposed MeLeMaD framework not only enhances malware detection capabilities but also provides a foundation for future research in adaptive cybersecurity solutions. By addressing the challenges posed by evolving threats in the cybersecurity landscape, MeLeMaD represents a significant advancement in malware detection technology. The MeLeMaD's adaptability, efficiency, and high performance make it a valuable contribution to the field of malware detection, paving the way for more robust and scalable solutions in the ever-evolving landscape of cybersecurity. Future research will focus on enhancing interpretability using explainable AI (XAI) techniques, improving detection of zero-day attacks through few-shot learning and continuous learning paradigms, and exploring hybrid approaches that combine static and dynamic analysis for more comprehensive detection.

\section*{Conflict of Interest}
The authors have no conflicts of interest to declare.




\bibliographystyle{elsarticle-num-names} 
\bibliography{references}

\end{document}